\documentclass[journal]{IEEEtran}

\usepackage[cmex10]{amsmath}
\usepackage{cite}
\usepackage{url}
\usepackage{color}

\usepackage{amssymb}
\usepackage{path}
\usepackage{verbatim}
\usepackage{algorithm}
\usepackage{framed}
\usepackage{algpseudocode}

\usepackage{color}
\usepackage{amsmath}
\usepackage{cases}
\usepackage{theorem}
\usepackage{subfig}
\usepackage{subfloat}

\usepackage{graphicx}

\usepackage{pgfplots}
\usepackage{pgfplotstable}
\pgfplotsset{compat=newest}

\newcommand{\R}{\ensuremath{\mathbb{R}}}
\newcommand{\N}{\ensuremath{\mathbb{N}}}
\newcommand{\spann}{\ensuremath{\operatorname{span}}}

\newcommand{\dist}{\ensuremath{\operatorname{dist}}}

\newtheorem{theorem}{Theorem}
{ \theorembodyfont{\normalfont} %\theorembodyfont{\rmfamily}

}

\newtheorem{lemma}[theorem]{Lemma}

\newcounter{enumctr}
\newenvironment{enum}{\begin{list}{(\roman{enumctr})}{\usecounter{enumctr}}}{\end{list}}

\DeclareFontFamily{U}{mathx}{\hyphenchar\font45}
\DeclareFontShape{U}{mathx}{m}{n}{<-> mathx10}{}
\DeclareSymbolFont{mathx}{U}{mathx}{m}{n}
\DeclareMathAccent{\widebar}{0}{mathx}{"73}

\begin{document}

\title{A Distributed and Privacy-Aware Speed Advisory System for Optimising Conventional and Electric Vehicles Networks}

\author{Mingming~Liu, Rodrigo H. Ord\'o\~nez-Hurtado, Fabian~Wirth, Yingqi Gu, Emanuele Crisostomi and Robert~Shorten % <-this % stops a space

\thanks{M. Liu and R. Ord\'o\~nez-Hurtado, joint first authors, are with the Hamilton Institute, Maynooth University, Maynooth, Ireland (e-mail: mliu@eeng.nuim.ie; Rodrigo.Ordonez@nuim.ie).}
\thanks{F. Wirth is with the Faculty of Computer Science and Mathematics,
  University of Passau, Germany (e-mail: fabian.lastname@uni-passau.de). }
%mistake in Fabian's email is intentional, please do NOT correct.
\thanks{Y. Gu, M. Liu, R. Ord\'o\~nez-Hurtado, R. Shorten are with the School of Electrical, Electronic and Communications Engineering, University College Dublin, Ireland (e-mail: yingqi.gu@ucdconnect.ie) }
\thanks{E. Crisostomi is with the Department of Energy, Systems, Territory and Constructions Engineering, University of Pisa, Italy (e-mail: emanuele.crisostomi@unipi.it). }
\thanks{R. Shorten is also with IBM Research, Dublin, Ireland (e-mail: ROBSHORT@ie.ibm.com).}% <-this % stops a space
}

\maketitle

%\textcolor[rgb]{1,0,0}{

\begin{abstract}
One of the key ideas to make Intelligent Transportation Systems (ITS) work effectively is to deploy advanced communication and cooperative control technologies among the vehicles and road infrastructures. In this spirit, we propose a consensus-based distributed speed advisory system that optimally determines a recommended common speed for a given area in order that the group emissions, or group battery consumptions, are minimised. Our algorithms achieve this in a privacy-aware manner; namely, individual vehicles do not reveal in-vehicle information to other vehicles or to infrastructure. A mobility  simulator is used to illustrate the efficacy of the algorithm, and hardware-in-the-loop tests involving a real vehicle are given to illustrate user acceptability and ease of the deployment.

\end{abstract}


\IEEEpeerreviewmaketitle

%%%%%%%%%%%%%%%%%%%%%%%%%%%%%%%%%%%%%%%%%%%%%%%%%%%
%%%%%%%%%%%%%%%%%%%%%%%%%%%%%%%%%%%%%%%%%%%%%%%%%%%
%%%%%%%%%%%%%%%%%%%%%%%%%%%%%%%%%%%%%%%%%%%%%%%%%%%
\section{Introduction}
%%%%%%%%%%%%%%%%%%%%%%%%%%%%%%%%%%%%%%%%%%%%%%%%%%%
%%%%%%%%%%%%%%%%%%%%%%%%%%%%%%%%%%%%%%%%%%%%%%%%%%%
%%%%%%%%%%%%%%%%%%%%%%%%%%%%%%%%%%%%%%%%%%%%%%%%%%%

At present, Intelligent Speed Advisory (ISA) systems, as a part of Advanced Driver Assistance Systems (ADASs), have become a fundamental part of Intelligent Transportation Systems (ITS). Such systems offer many potential benefits, including improved vehicle and pedestrian safety, better utilisation of the road network, and reduced emissions. Recently, many papers have appeared on this topic reflecting the problem from the viewpoint of road operators, infrastructure providers, and transportation solution providers \cite{VANETenter,VANETsafe1,VANETsafe2,carsten2008,adell2008,hounsell2009,tradisauskas2009,UCTCPolicy}.\newline

In this paper, we consider the design of a speed advisory system (SAS) making use of vehicle-to-vehicle/infrastructure (V2X) technologies. Our starting point is the observation that different vehicle classes are designed to operate optimally at different vehicle speeds and at different loading conditions. Thus, a recommended speed, or speed limit, may be optimal for one vehicle and not for others. One possible way to handle this is by recommending a different speed to different vehicles to take into account their differences (e.g., vehicle type, vehicle age, fuel mode, load, desired time of arrival).
However, asking different cars to drive at different speeds does not make sense in any practical scenario. On the other hand, groups of cars following a common speed occurs frequently in practice. For example, roughly speaking, in highway driving cars tend to follow a given speed limit where possible. This tendency is increased in situations where intelligent tempomats are deployed. Other situations where common cars follow common speeds include dedicated lanes, and special zones in cities. Finally, generic benefits of following a common speed include, reduced emissions (due to less frequent accelerations/decelerations), reduced energy consumption, increased throughput, and increased safety \cite{Nature,ICCVE2013,Nature2,SUMO_Optimal}.\newline

In this paper, we consider the problem of recommending a common speed to a set of cars. Our ISA will be constructed with an instantaneous optimality goal in mind. For example, we may wish to minimise the instantaneous group emissions, or the instantaneous group energy consumption. Specifically, we are interested in addressing the following questions:
\begin{itemize}
\item[(i)]
Firstly, we are interested in designing an ISA system to recommend the same speed to a network of conventional vehicles (i.e., with an Internal Combustion Engine (ICE)) travelling along an extra-urban route (e.g., a highway), connected via a V2X communication system, such that the total emission is minimised;
\item[(ii)]
Secondly, we are interested in designing an ISA system to recommend the same speed to a network of EVs travelling in an urban network (e.g., the city centre), connected via a V2X communication system, such that the total battery consumption is minimised.\newline
\end{itemize}

We shall show that these problems can be solved in a privacy preserving manner without a large communication burden on the vehicle-to-vehicle (V2V) network.
Furthermore, we believe that algorithms addressing these problems will be of use when
cars are asked to follow a constant speed to achieve certain goals.
These include, transit through
environmental and safety zones, highway driving with adaptive speed-limits
(up to some maximum deviation from the nominal speed limit), and
eco-driving for fleets of autonomous vehicles travelling in special lanes.\newline

This paper is an extended version of the idea presented in \cite{ICCVE2014}. In order to allow vehicles to collaboratively compute the optimal recommended speed, we shall further assume they are equipped with V2X technologies, can exchange information with their neighbours, and can exchange limited information with the infrastructure. We shall show that one can design, using very simple ideas, an effective SAS in a manner that preserves the privacy of individual vehicles. Extensive simulations, including hardware-in-the-loop (HIL) tests using real vehicles, are given to demonstrate the efficacy of our approach.

%%%%%%%%%%%%%%%%%%%%%%%%%%%%%%%%%%%%%%%%%%%%%%%%%%%
%%%%%%%%%%%%%%%%%%%%%%%%%%%%%%%%%%%%%%%%%%%%%%%%%%%
%%%%%%%%%%%%%%%%%%%%%%%%%%%%%%%%%%%%%%%%%%%%%%%%%%%
\section{Related work}
%%%%%%%%%%%%%%%%%%%%%%%%%%%%%%%%%%%%%%%%%%%%%%%%%%%
%%%%%%%%%%%%%%%%%%%%%%%%%%%%%%%%%%%%%%%%%%%%%%%%%%%
%%%%%%%%%%%%%%%%%%%%%%%%%%%%%%%%%%%%%%%%%%%%%%%%%%%

In this section, we give a brief review of some related work.
First note that a detailed review of this topic is given in \cite{Ordonez-Hurtado2014}.
Conventional systems are described in \cite{adell2008,hounsell2009,tradisauskas2009,gallen2013,Murray2007}. These papers describe various aspects of the ISA design process, which includes the design of driver display systems, the incorporation of external environmental information, and the algorithmic aspects of speed and distance recommendations. Recently, there has been a strong trend to also include traffic density information; references \cite{Wang2010,Garelli2011,Tyagi2012,Schakel2013,Ordonez-Hurtado2013,Ordonez-Hurtado2014}
describe work in this direction. In these works, density information is included in the procedure via loop detectors or via explicit density estimation using V2V technology. The differentiating feature of the approach followed in this paper is that density and composition of the vehicle fleet is also used, but in an implicit manner as part of the optimisation algorithm. Finally, we note that there is a huge body of work on cooperative control of vehicles and its connection to consensus algorithms \cite{Kato2002,Choy2003,Murray2007,Flemisch2008}.
It is important to note that we are designing a SAS and not a cooperative control system. This distinction is important as it allows us to ignore string stability effects which are a fundamental limitation of many cooperative control architectures \cite{Hedrick1994,Swaroop1996,Klinge2009,Knorn2013}.

%%%%%%%%%%%%%%%%%%%%%%%%%%%%%%%%%%%%%%%%%%%%%%%%%%%
%%%%%%%%%%%%%%%%%%%%%%%%%%%%%%%%%%%%%%%%%%%%%%%%%%%
%%%%%%%%%%%%%%%%%%%%%%%%%%%%%%%%%%%%%%%%%%%%%%%%%%%
\section{Model and Algorithm}\label{s:Model_Algorithm}
%%%%%%%%%%%%%%%%%%%%%%%%%%%%%%%%%%%%%%%%%%%%%%%%%%%
%%%%%%%%%%%%%%%%%%%%%%%%%%%%%%%%%%%%%%%%%%%%%%%%%%%
%%%%%%%%%%%%%%%%%%%%%%%%%%%%%%%%%%%%%%%%%%%%%%%%%%%

%-------------------------------------------------------------------------------------------------------------------------------------------------------------------------------------------
\subsection{Problem Statement}\label{s:Problem_Statement}
%-------------------------------------------------------------------------------------------------------------------------------------------------------------------------------------------

%We consider a scenario in which a number of vehicles are driving along a given stretch of a highway with several lanes in the same direction. We wish to find a common recommended speed for these vehicles that takes into account the composition of the vehicles (individual vehicle types) and the number of vehicles. Note that the assumption on different lanes of the highway allows vehicles to overtake whenever it is appropriate. Let $N$ denote the total number of vehicles on a particular section of the highway where the ISA broadcast signal can be received. Each vehicle is equipped with a specific communication device (e.g. a mobile phone with access to WiFi/3G networks) so that it is able to receive/transmit messages from/to either nearby vehicles or available road infrastructure (e.g. a base station). We assume that each vehicle can communicate a limited amount of information with the infrastructure, and that the infrastructure can broadcast information to the entire network of cars, and each vehicle can send a broadcast signal to its neighbours. \newline

Our objective is to develop an algorithm that converges to a common recommended speed for a fleet of vehicles driving along the same stretch of a road (e.g., a highway), or in the same area (e.g., in the city centre). Let $N$ denote the total number of vehicles in the fleet, and let us assume that they all are able to receive the ISA broadcast signal. For this purpose, each vehicle is equipped with a specific communication device (e.g., a mobile phone with access to WiFi/3G networks) so that they are able to receive/transmit messages from/to either nearby vehicles or available road infrastructure (e.g. a base station). We assume that each vehicle can communicate a limited amount of information with the infrastructure, the infrastructure can broadcast information to the entire network of cars, and each vehicle can send a broadcast signal to its neighbours. \newline

For convenience, we assume that all vehicles have access to a common clock (for example, a GPS clock). Let $k\in\left\{1,2,3,...\right\}$ be a discrete-time instant in which new information from vehicles is collected and new speed recommendations are made. Let $s_i\left(k\right)$ be the recommended speed of the vehicle $i\in\underbar{N}:=\left\{1,2,...,N\right\}$ calculated at time instant $k$. Thus, the vector of recommended speeds for all vehicles is given by $\textbf{s}\left(k\right)^{\textrm{T}}: = \left[s_1\left(k\right), s_2\left(k\right),...,s_N\left(k\right)\right]$, where the superscript $\textrm{T}$ represents the transposition of the vector. Note that between two consecutive time instants $(k, k+1)$, the recommended speeds are constant while the driving speeds are time-varying real-valued variables. We denote by $N_k^i$ the set of neighbours of vehicle $i$ at time instant $k$, i.e., those vehicles which can successfully broadcast their recommended speeds to vehicle $i$. \newline

\begin{figure}[h]
	\begin{center}
		{\includegraphics[width=2.8in]{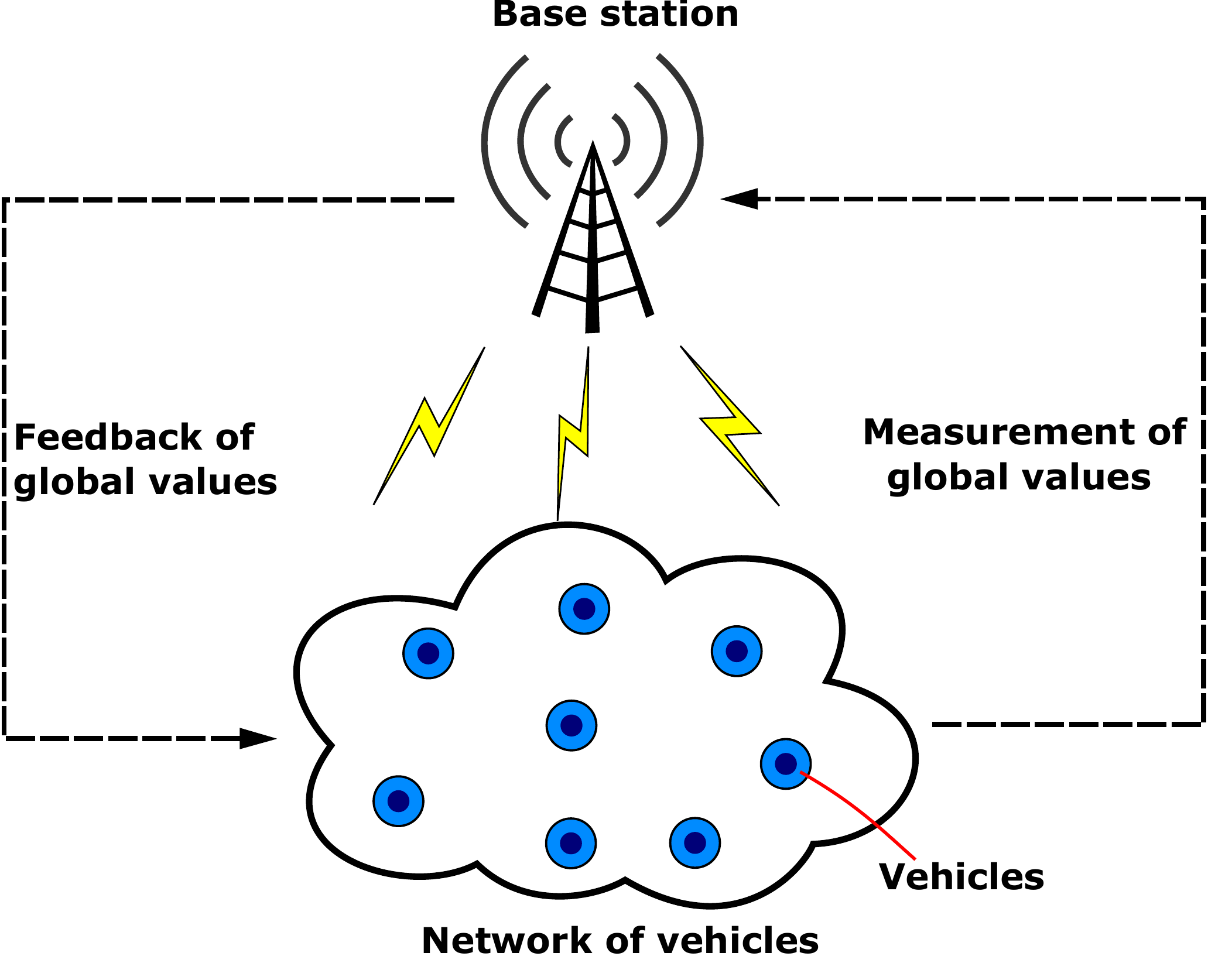}}
		\caption{Schematic diagram of the framework \cite{TAC}.}
		\label{fig1:schematic}
	\end{center}
\end{figure}

In addition, we assume that each vehicle $i$ can evaluate a function $f_i$ that determines its average emissions, in the ICE case study, or its average energy consumption in the EV case, were it to be travelling at the recommended speed $s_i\left(k\right)$. Such functions are typically convex functions of the vehicle speed, as it will be shown in more detail in the following sections. We shall further assume that these functions are continuously differentiable and with a Lipschitz continuous first derivative $f'_i$ which is assumed with positive bounded growth rate, i.e.,
\begin{equation} \label{bdd}
0 < d_{\min}^{\left(i\right)} \leq \frac{f'_{i}\left(a\right) - f'_{i}(b) }{a - b} \leq d_{\max}^{\left(i\right)},
\end{equation}
for all $a,b\in \R$ such that $a$${\small\neq}$$b$, and suitable positive constants $d_{\min}^{\left(i\right)}$, $d_{\max}^{\left(i\right)}$.
A schematic diagram of the above is illustrated in Fig. \ref{fig1:schematic}.
In this context, we consider the following problem. \newline

\noindent \textbf{Problem 1:} {\it Design an ISA system for a network of vehicles connected via V2X communication systems, in order to recommend a common speed that minimises the amount of emissions, or the total energy consumption, of the whole fleet of vehicles.} \newline

The optimisation problem that needs to be solved in order to address Problem 1 can be formulated as follows:
\begin{equation} \label{eq:opt}
\begin{gathered}
\underset{\textbf{s} \in \R^N}{\min} \quad
\sum\limits_{j\in\underbar{N}} f_{j}\left(s_j \right),\\
{\text{s.t.}} ~
s_i = s_j, ~ \forall i \neq j \in \underbar{N}.
\end{gathered}
\end{equation}

The above is an optimised consensus problem and can be solved in a variety of ways (for example using ADMM \cite{boydADMM1,ADMM2,ADMM3}). Our focus in this present work is not to construct a fully distributed solution to this problem, but rather to construct a partially distributed solution which allows rapid convergence to the optimum, without requiring the vehicles to exchange information that reveals individual cost functions to other vehicles. This is the privacy preserving component of our problem statement.\newline

{\noindent \bf Comment:} Note that \eqref{eq:opt} involves the recommended speeds $s_j$ and not the actual speeds at which the vehicles are travelling in reality. This implies, among other things, that the optimal solution of (\ref{eq:opt}) is the same, disregarding whether vehicles follow in reality to the recommended speed or not.
\newline

{\noindent \bf Comment:} In the following, we shall not solve \eqref{eq:opt} in a single step.  Instead, we shall propose an iterative algorithm that at each step yields individual recommended speeds that will eventually converge to the same value on the consensus constraints. \newline
%Thus, our objective for the minimisation problem \eqref{eq:opt} is to seek for the optimal solution of the recommended speeds under a consensus constraint. Clearly the constraints for vehicles to travel at roughly the same speed on a section of highway is a reasonable practical assumption.\newline

To solve (\ref{eq:opt}) we use the iterative feedback scheme
\begin{equation} \label{eq:fw-tvsys0}
\begin{gathered}
\textbf{s}\left(k+1\right) = P\left(k\right)\textbf{s}\left(k\right) + G\left(\textbf{s}\left(k\right)\right)e,\\
\end{gathered}
\end{equation}
where $\left\{ P\left(k\right) \right\} \in \mathbb{R}^{N \times N}$ is a sequence of row-stochastic matrices\footnote{Square matrices with non-negative real entries, and rows summing to $1$.}, $e\in \mathbb{R}^{N}$ is a column vector with all entries equal to 1, and $G:
\R^{N} \mapsto \R$ is a continuous function with some assumptions to satisfy as we shall see in Theorem \ref{t:globalstab}. Algorithms of this type were proposed and studied among others in \cite{Knorn,knornIJC1,knornIJC2}; the principal theoretical contribution here is to extend this framework to a new class of optimisation problems and to give conditions guaranteeing their convergence.\newline

\noindent \textbf{Remark:} Eq. (\ref{eq:fw-tvsys0}) describes a consensus algorithm with an input. In the first part of the right-hand side of (\ref{eq:fw-tvsys0}), $P(k)$ creates pressure on $\textbf{s}(k)$ so that all components of $\textbf{s}(k)$ achieve a common value. The second part of (\ref{eq:fw-tvsys0}) describes the regulation constraint that the nonlinear function of $\textbf{s}(k)$ must satisfy. Thus, we can use Eq. (\ref{eq:fw-tvsys0}) to force all components of $\textbf{s}(k)$ to achieve a common value and at the same time to satisfy a nonlinear constraint. The choice of $G$ thus determines which constraint is satisfied when $\textbf{s}(k)$ reaches the equilibrium. In principle, many choices of $G$ can be used. In this paper, we choose $G$ to solve an optimisation problem. The principal mathematical considerations in deploying this algorithm are to determine the mathematical conditions on $G$ so that (\ref{eq:fw-tvsys0}) actually converges to the desired value. While this is not a primary concern of this paper, rather we are interested in the applications of this, in the remainder of this section we shall give a brief overview of such considerations.\newline

We will require that (\ref{eq:opt}) has a unique solution.  Note that, it follows from elementary optimisation theory that if all the $f_i$'s are strictly convex functions, then the optimisation problem (\ref{eq:opt}) has a solution if and only if there exists a $y^{*} \in \R$ satisfying
\begin{equation} \label{optcondition}
\sum\limits_{j=1}^{N} f'_j(y^{*}) = 0.
\end{equation}
In this case, $y^*$ is unique by strict convexity, and the unique optimal point of (\ref{eq:opt}) is given by 
\begin{equation}
    \label{eq:1}
\textbf{s}^{*} := y^{*}e \in \R^N.    
\end{equation}
In order to obtain convergence of \eqref{eq:fw-tvsys0} we select a feedback signal
\begin{equation}
    \label{eq:Gdef}
 G\left(\textbf{s}\left(k\right)\right) = -\mu \sum_{j=1}^{N} f'_j\left(s_j\left(k\right)\right),
\end{equation}
and we obtain the dynamical system
\begin{equation} \label{eq:iter}
\textbf{s}(k+1) = P(k)\textbf{s}(k) - \mu  \sum\limits_{j=1}^{N} f'_j(s_j(k))e,\,\,\,\,\mu \in \R.
\end{equation}

In \cite{TAC} it is shown that if $\left\lbrace P(k) \right\rbrace_{k \in
  \mathbb{N}}$ is a uniformly strongly ergodic sequence\footnote{That is,
  for every $k_0 \in \N$ the sequence $P(k_0)$, $P(k_0+1)P(k_0)$,
  $\ldots$, $P(k_0+ \ell)\cdots P(k_0), \ldots$ converges to a rank one
  matrix. See \cite{TAC} for further details.} and $\mu$ is chosen according to 
\begin{equation} \label{rangemu}
0 < \mu < 2 \left(\sum\limits_{j=1}^{N} d_{\textrm{max}}^{(j)}\right)^{-1},
\end{equation} 
then (\ref{eq:iter}) is uniformly globally asymptotically stable at the unique optimal point $\textbf{s}^{*} = y^{*}e$ of (\ref{eq:opt}). 
%For completeness, we formally state this as a theorem. 
%In Theorem \ref{t:globalstab}, we refer to a one-dimensional Lure system associated to (\ref{eq:fw-tvsys0}). This new system is used
%to demonstrate the stability of system (\ref{eq:fw-tvsys0}), for which the form of the Lure system is given by
Such systems were studied in \cite{Knorn} and formally analysed in \cite{TAC}.  
For completeness, we formally state the relevant results from these works as a theorem  (Theorem 1).
An overview of the proof is given in the appendix, and the interested reader
may refer to \cite{TAC} for details.

\begin{theorem}[\cite{TAC}]
\label{thm:thm01}
\label{t:globalstab}
Consider the optimisation problem \eqref{eq:opt} and the optimisation
algorithm \eqref{eq:fw-tvsys0}, and the associated Lure system
\begin{equation}
\label{eq:fw-tvsys0_1dim}
\begin{array}{rcl}
y\left(k+1\right) &=& h\left(y\left(k\right)\right),\\
h\left(y\right) &:=&  y + G\left(ye\right)\,.
\end{array}
\end{equation}
If $G$ is defined by \eqref{eq:Gdef} and the
condition \eqref{rangemu} holds, then the following assertions hold:
\begin{enum}
  \item There exists a unique, globally
    asymptotically stable fixed point $y^* \in \R$ of the
    Lure system \eqref{eq:fw-tvsys0_1dim}.
  \item The fixed point $y^*$ of (i) satisfies the optimality condition
    \eqref{optcondition}, and thus $y^* e \in \R^N$ is the unique optimal
    point for the optimisation problem \eqref{eq:opt}.
  \item If, in addition, $\left\{ P\left(k\right) \right\}_{k\in\N}
    \subset \R^{N \times N}$ is a strongly ergodic sequence of
    row-stochastic matrices, then $y^* e$ is a globally asymptotically
    stable fixed point for system \eqref{eq:fw-tvsys0}.
\end{enum}
\end{theorem}

An outline of the proof can be found in Appendix \ref{Proofoutline}.
To apply the Theorem 1 to solve the optimisation problem we proceed as
follows. For each $k$ we define the $P\left(k\right)$ as 
\begin{equation} \label{Pk}
P_{i,j}\left(k\right)=\left\{ \begin{array}{cc}
1-\sum_{j\in N_{k}^{i}}\eta_j, & \mbox{if }j=i,\\
\eta_j, & \mbox{if }j\in N_{k}^{i},\\
0, & \mbox{otherwise.}
\end{array}\right.,
\end{equation}
where $i,j$ are the entries' indexes of the matrix $P\left(k\right)$, and $\eta_j \in \mathbb{R}$ is a weighting factor. 
For example, a convenient choice $\eta_j$ is $\frac{1}{\left| N_k^{i}
  \right| + 1} \in \left(0,\frac{1}{N-1}\right)$, where $\left| \bullet
\right|$ denotes cardinality, giving rise to an equal weight factor for
all elements in the reference speed vector
$\textbf{s}\left(k\right)$.\newline

The assumption of uniform strong ergodicity holds if the neighborhood graph associated with the problem has suitable connectedness properties. If sufficiently many cars travel on a given area, it is reasonable to expect that this graph is strongly connected at most time instances. Weaker assumptions are possible but we do not discuss them here for reasons of space; see \cite{Liter3} for possible assumptions in this context.\newline

Now, we propose the Optimal Decentralised Consensus Algorithm for solving (\ref{eq:opt}) as shown in Algorithm \ref{alg:Algorithm1}. The underlying assumption here is that at all time instants all cars communicate their value $f'_{j}\left(s_j\left(k\right)\right)$ to the base station, which reports the aggregate sum back to all cars. This is precisely the privacy preserving aspect of the algorithm, as cars do not have to reveal their cost functions to anyone. Also implicit information as derivatives of the cost function at certain speeds is only revealed to the base station but not to any other agent involved in the system.

\begin{algorithm}[htbp]
	\caption{Optimal Decentralised Consensus Algorithm}
	\begin{algorithmic}[1]
		\For{$k=1,2,3,..$}
			\For{each $i \in \underbar{N}$}
			\State Get $\tilde{F}\left(k\right) = \sum\limits_{j\in\underbar{N}} f'_{j}\left(s_j\left(k\right)\right)$ from the base station.
			\State Get $s_j\left(k\right)$ from all neighbours $j\in N_k^{i}$.
			\State Do $q_i\left(k\right)=\eta_i\cdot \sum\limits_{j\in N_k^{i}}\left(s_j\left(k\right)-s_i\left(k\right)\right)$.
			\State Do $s_i\left(k+1\right) = s_i\left(k\right) + q_i\left(k\right) - \mu \cdot \tilde{F}\left(k\right)$.
			\EndFor
		\EndFor
	\end{algorithmic}
	\label{alg:Algorithm1}
\end{algorithm}

{\noindent \bf Comment:} We note that in any practical implementation of the previous algorithm, the recommended speed may be bounded above and below by the road operator (i.e., should the optimal solution computed by Algorithm \ref{alg:Algorithm1} exceed speed limits, or be unreasonably low).

%%%%%%%%%%%%%%%%%%%%%%%%%%%%%%%%%%%%%%%%%%%%%%%%%%%
%%%%%%%%%%%%%%%%%%%%%%%%%%%%%%%%%%%%%%%%%%%%%%%%%%%
%%%%%%%%%%%%%%%%%%%%%%%%%%%%%%%%%%%%%%%%%%%%%%%%%%%
\section{A Speed Advisory System for Conventional Vehicles} \label{SUMO}
%%%%%%%%%%%%%%%%%%%%%%%%%%%%%%%%%%%%%%%%%%%%%%%%%%%
%%%%%%%%%%%%%%%%%%%%%%%%%%%%%%%%%%%%%%%%%%%%%%%%%%%
%%%%%%%%%%%%%%%%%%%%%%%%%%%%%%%%%%%%%%%%%%%%%%%%%%%

In this section, we evaluate the performance of Algorithm 1 using SUMO \cite{Behrisch2011}. SUMO is a microscopic, open source road traffic simulator, which is frequently used for validation and prediction purposes in the ITS community. First, we evaluate it using conventional simulations, and then use a real vehicle embedded into a hardware-in-the-loop (HIL) emulation. In particular, we perform the following experimental activities:
\begin{enumerate}
\item First, we compare the pollution emission of a fixed fleet of vehicles (with initial non-optimal speeds) before and after Algorithm 1 is turned on.
\item We then provide simulation results which are dynamic in nature, where vehicles enter and leave a section of the highway where the SAS is active; in doing so, we provide a discussion about the rate of convergence of Algorithm 1, and also evaluate what happens when some vehicles decide not to follow the recommended speed.
\item Finally, we give a HIL emulation with a real vehicle travelling on a real road, embedded into a emulated network with a fixed number of simulated vehicles.
\newline
\end{enumerate}

In all the above experiments, we evaluate the performance of Algorithm 1 for $\eta = 0.001$ and $\mu = 0.01$ (a convergent set-up), and all the vehicles are also subject to the driving constraints arising from the SUMO simulation (e.g., in terms of acceleration/deceleration profiles).

\subsection{Cost funtions to represent pollution emission in ICE vehicles}
\begin{figure}[h] 
	\centering
	\includegraphics[width=3.3in]{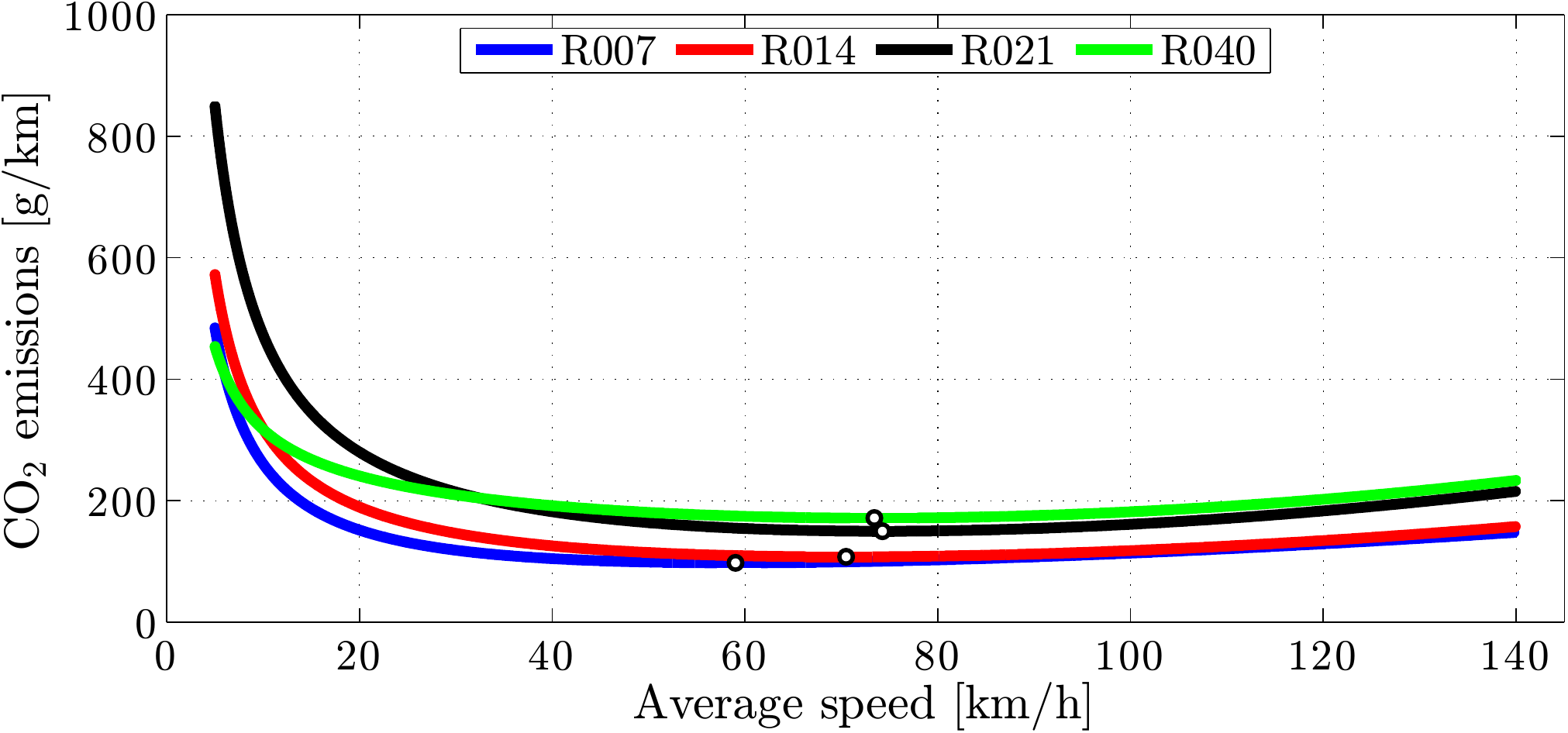}
	\caption{Curves for the CO$_{\mbox{\footnotesize{2}}}$ emission types in Table \ref{tab:Table_emissions}. Black circles mark the minimum point on each curve.}
	\label{figco2}
\end{figure}

%The idea in all situations is to show the benefits of Algorithm 1. For this purpose,
We shall adopt the average-speed model proposed in \cite{emfactor} to model each cost function $f_{i}$ in function of the average speed $s$ as
\begin{equation}
\begin{gathered}
f_{i}=
\text{k}\left(\frac{\text{a} + \text{b}s_{i}%\left(k\right)
+ \text{c}s^{2}%\left(k\right)
+ \text{d}s^{3}%\left(k\right)
+ \text{e}s^{4}%\left(k\right)
+ \text{f}s^{5}%\left(k\right)
+ \text{g}s^{6}%\left(k\right)
}
{s%\left(k\right)
}\right),
\end{gathered}
\end{equation}
where $\text{a},\text{b},\text{c},\text{d},\text{e},\text{f},\text{g},\text{k} \in \mathbb{R}$ are used to specify different levels of emissions by different classes of vehicles. In addition, we use the emission types from \cite{emfactor} shown in Table \ref{tab:Table_emissions} and Fig. \ref{figco2}, corresponding to petrol cars/minibuses with up to 2.5 tons of gross vehicle mass. Also, we use the following vehicle types:
\begin{itemize}
	\item Type 1: accel. 2.15 m/s$^{2}$,
	decel. 5.5 m/s$^{2}$, length 4.54 m.
	\item Type 2: accel. 1.22 m/s$^{2}$,
	decel. 5.0 m/s$^{2}$, length 4.51 m.
	\item Type 3: accel. 1.75 m/s$^{2}$,
	decel. 6.1 m/s$^{2}$, length 4.45 m.
	\item Type 4: accel. 2.45 m/s$^{2}$,
	decel. 6.1 m/s$^{2}$, length 4.48 m.
\end{itemize}

\begin{table}[h]
\caption{Emission factors for some CO$_{\mbox{\footnotesize{2}}}$ emission types reported in \cite{emfactor}. Here, $\left\{\text{e},\text{f},\text{g}\right\}=0$ and $\text{k}=1$.}
\begin{centering}
\begin{tabular}{|c|c|c|c|c|}
\hline 
Type & a & b & c & d \tabularnewline
\hline 
\hline 
R007 & { 2.2606E+3} & { 3.1583E+1} & { 2.9263E-1} & { 3.0199E-3} \tabularnewline
\hline 
R014 & { 2.5324E+3} & { 6.8842E+1} & { -4.3167E-1} & { 6.6776E-3} \tabularnewline
\hline 
R021 & { 3.7473E+3 } & { 1.0571E+2} & { -8.5270E-1} & { 1.0318E-2} \tabularnewline
\hline 
R040 & { 1.2988E+3} & { 2.0203E+2} & { -1.5597E-0} & { 1.2264E-2} \tabularnewline
\hline  
\end{tabular}
\par\end{centering}
\label{tab:Table_emissions}
\end{table}

%-------------------------------------------------------------------------------------------------------------------------------------------------------------------------------------------
\subsection{\label{sub:SUMO-simulation-static}SUMO Simulations with a Fixed Number of Vehicles}
%-------------------------------------------------------------------------------------------------------------------------------------------------------------------------------------------

In this experiment we consider 40 vehicles travelling along a highway. The set-up for this set of experiments is as follows.
\begin{itemize}
	\item Road: A straight 5 km long highway with 4 lanes.
	\item Duration of the simulation: 600 s.
	\item Algorithm sampling interval: $\Delta T $ = 1s.
	\item Switch-on time: the algorithm is activated at time 300 s.\newline
\end{itemize}

%\textbf{Comment:} For comparison purposes, the assumption on the number and the initial speed distribution of vehicles are consistent with our previous work in \cite{ICCVE2014}.\newline

\begin{figure}[h]
\centering
\includegraphics[width=3.3in]{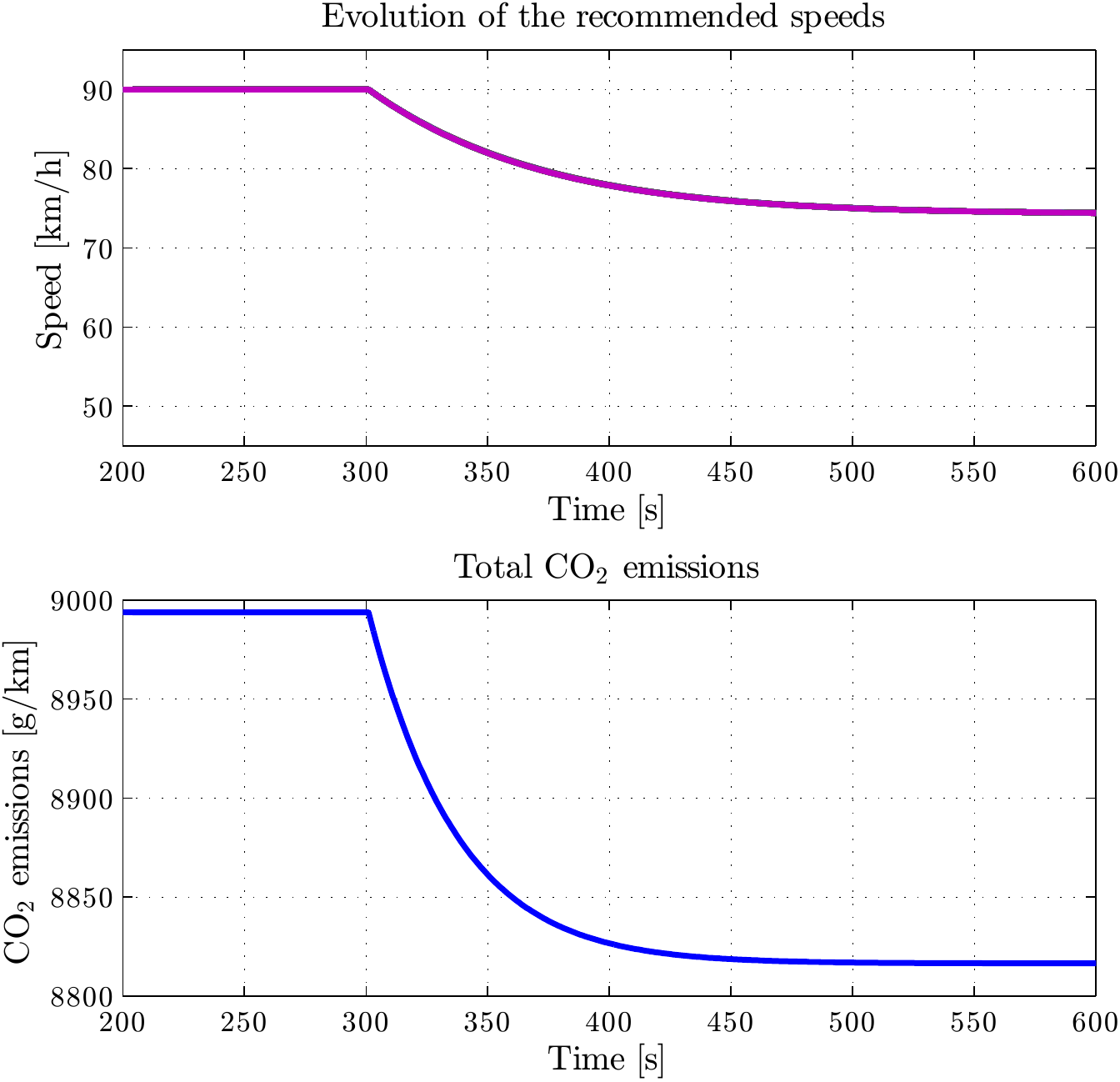}
\caption{Simulation results for the static case, before and after the activation of the algorithm at time step 300 s. Setup: 32 vehicles of emission type R007 and 8 vehicles of emission type R021, with uniform distribution of vehicle types.}
\label{New_Static_Fixed}
\end{figure}

\begin{figure}[h]
\centering
\includegraphics[width=3.3in]{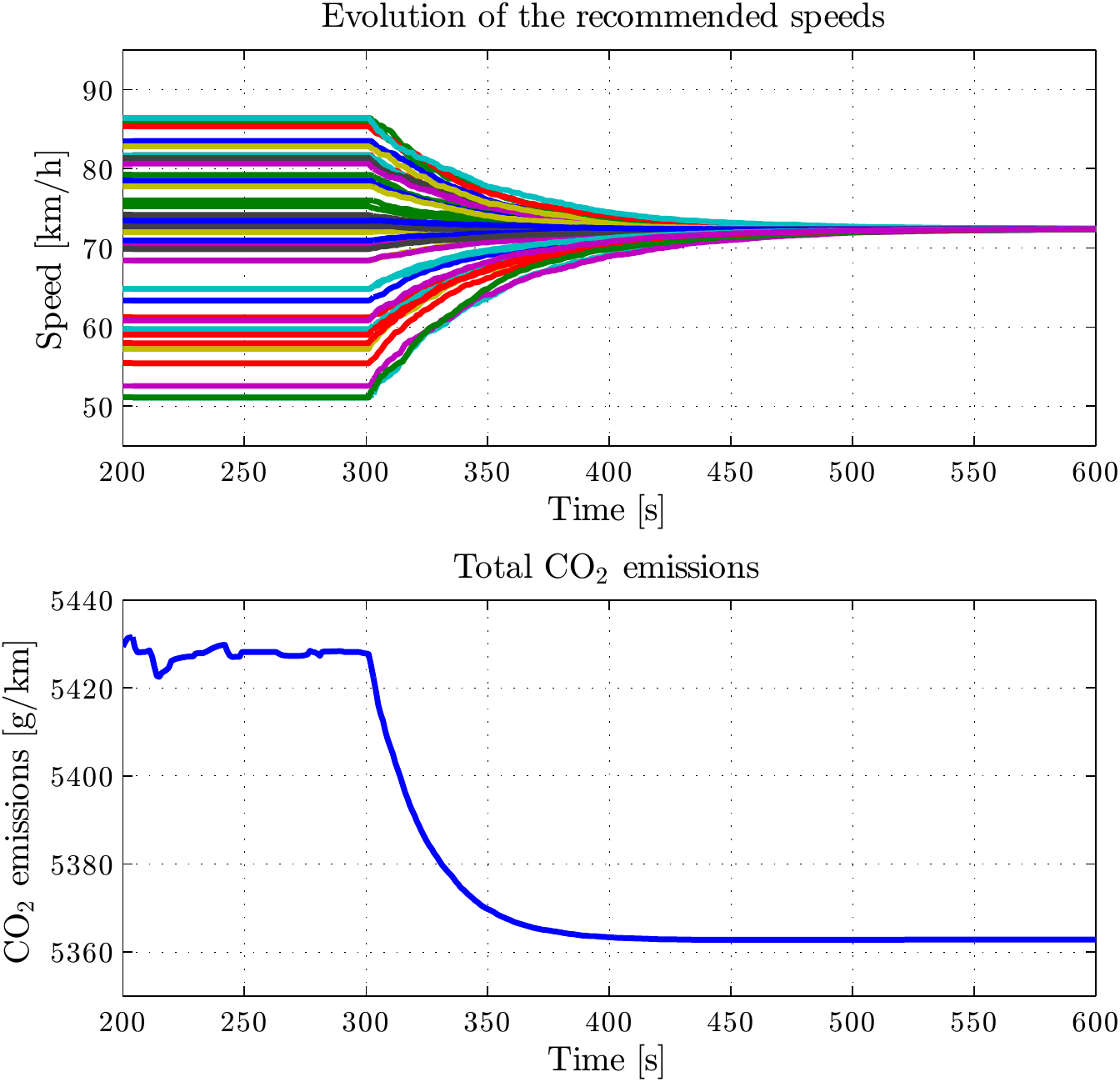}
\caption{Simulation results for the static case, before and after the activation of the algorithm at time step 300 s. Setup: uniform distribution of emission types in R014, R021 and R040, with uniform distribution of vehicle types.}
\label{New_Static_Range}
\end{figure}

The experimental results are given in Fig. \ref{New_Static_Fixed} and Fig. \ref{New_Static_Range}. As we can see from Fig. \ref{New_Static_Fixed}, making a small change in a common non-optimal initial speed yields a significant improvement in CO$_2$ emissions, as about 172 g of CO$_2$ are saved every kilometer. If this result is integrated over a whole road network, and integrated over a longer time horizon (e.g., a year), significant improvements can be observed. For instance, the savings are comparable to those that can be achieved if a EURO-3 vehicle (e.g., R039) is substituted by a EURO-4 vehicle (e.g., R040), where the savings are around 10 g/km.\newline

In the second experiment (see Fig. \ref{New_Static_Range}), we assume that the vehicles are initially driving at random speeds around the optimal one. Then, once the SAS is activated, the vehicles change their speed to eventually converge to the optimal one. Significant CO$_2$ savings can be noticed in this example again; however, carbon savings here are less than in the first experiment, since now some vehicles were already driving at (or very close to) the optimal speed.

%-------------------------------------------------------------------------------------------------------------------------------------------------------------------------------------------
\subsection{SUMO Simulations with a Dynamic Number of Vehicles} \label{threeB}
%-------------------------------------------------------------------------------------------------------------------------------------------------------------------------------------------

We now consider a dynamic scenario, with a time-varying number of vehicles. To do this we partition the highway into three consecutive sections L1, L2 and L3. We then proceed as follows. First, vehicles enter the uncontrolled section L1, with constant speed (randomly chosen in a given range). After completing L1, vehicles enter the section L2 where the SAS is always active, so the vehicles iteratively update their recommended speed using Algorithm 1. After completing L2, they finally enter section L3 where they travel freely. The experiments are setup as follows.
\begin{itemize}
	\item Road: three consecutive straight edges:
	\begin{itemize}
		\item L1: 5 km long highway with 4 lanes, uncontrolled;
		\item L2: 5 km long highway with 4 lanes, ISA controlled;
		\item L3: 5 km long highway with 4 lanes, uncontrolled.
	\end{itemize}
	
	\item Total number of cars: 650, with uniform distribution of both emission types among R014, R021 and R040, and vehicle types.
	
	\item Vehicular flow entering L1: one new car every 2 seconds until simulation time 1300 s.
	
	\item Length of simulation: 3010 s.
	
	\item Window size for the calculation of the moving average (MA) of CO$_2$ emissions for visualisation purposes: 500 time steps.
	
	\item Travelling speeds for cars on L1 are randomly chosen with uniform distribution in 3 scenarios:
	\begin{itemize}
		\item Case 1, constant speeds in $\left(80,100\right)$ km/h.
		\item Case 2, constant speeds in $\left(60,80\right)$ km/h.
		\item Case 3, constant speeds in $\left(40,60\right)$ km/h.\newline
	\end{itemize}
	
\end{itemize}

\begin{figure}[htbp]
	\begin{center}
		{\includegraphics[width=3.3in]{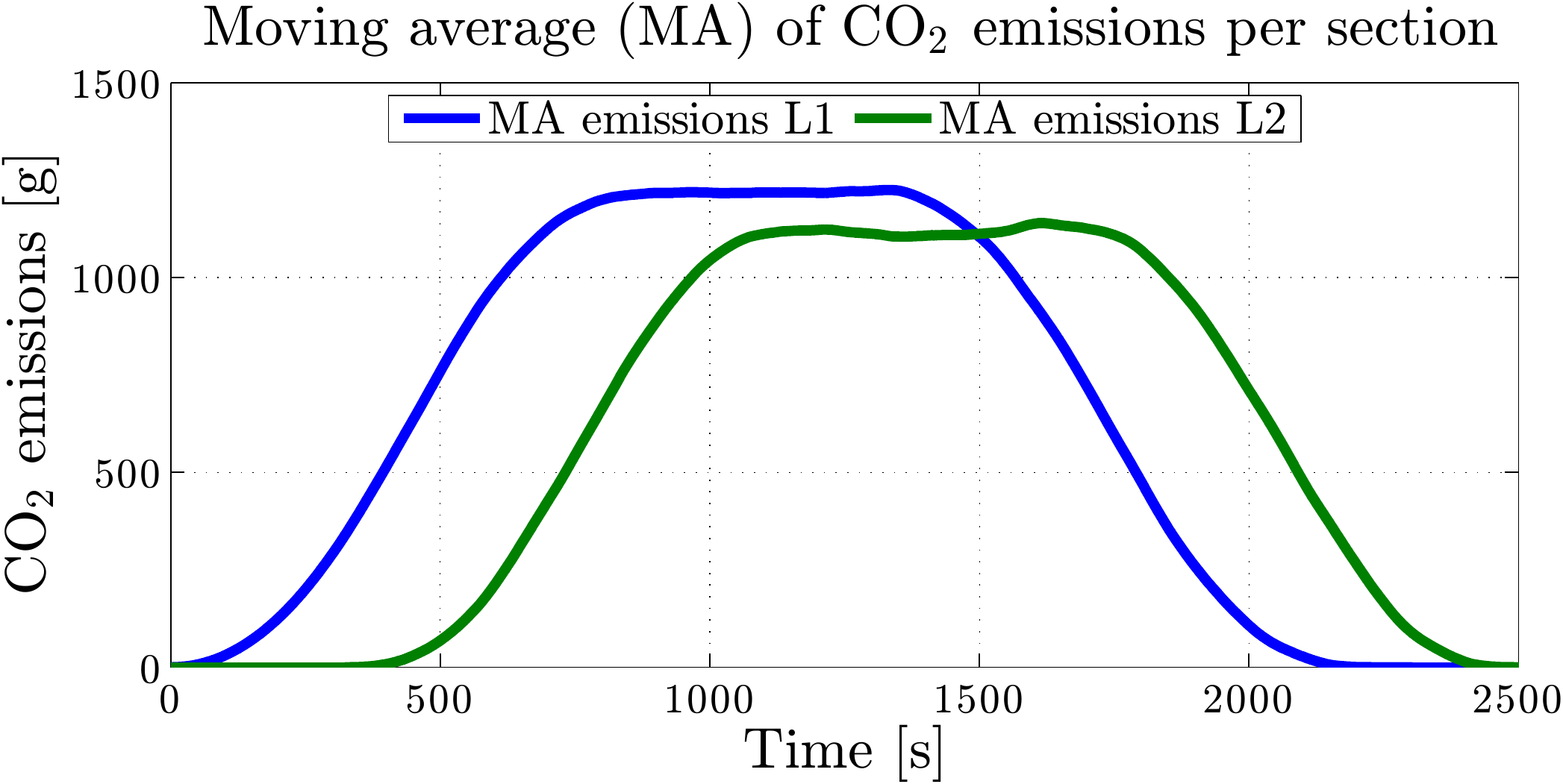}}
		\caption{Simulation results for the dynamic case: MA of CO$_2$ emissions on L1 and L2, for initial speeds on L1 in Case 3.}
		\label{fig:Case3}
	\end{center}
\end{figure}

Note that even though this is a dynamic situation, the vehicle density on each part of the road becomes almost constant after certain time. A sample of simulation results is given in Fig. \ref{fig:Case3}, which reveals what might be expected from the initial experiments. Namely, the further vehicles are away from the optimal speed, the more is to be gained by deploying the ISA.\newline

To generalise the previous results, we conducted one hundred random experiments for each of the three cases described above. In each experiment, we collected the simulation data of the total CO$_2$ emission generation on each section of the highway from SUMO. Table \ref{table2} summarises the aggregated results of this exercise, and clearly demonstrates the benefits of the ISA.\newline

\begin{table}[h!]
	\caption{General results: total emissions per lane.}
	\begin{center}
	\begin{tabular}{c|c|c|c|c|c|c|}
		\cline{2-5} 
		& \multicolumn{4}{c|}{Total Emissions$*$ {[}tonnes{]}} & \multicolumn{2}{c}{}\tabularnewline
		\cline{2-7} 
		& \multicolumn{2}{c|}{L1 (Uncontrolled)} & \multicolumn{2}{c|}{L2 (Controlled)} & \multicolumn{2}{c|}{Improvement}\tabularnewline
		\cline{1-7} 
		\multicolumn{1}{|c|}{Case} & Mean & $\sigma$ & Mean & $\sigma$ & Mean & $\sigma$ \tabularnewline
		\hline 
		\multicolumn{1}{|c|}{ 1} & 1.5199 & 0.00114 & 1.4684 & 0.0002 & 3.40\% & 0.07\tabularnewline
		\hline 
		\multicolumn{1}{|c|}{ 2} & 1.4751 & 0.0004 & 1.4648 & 0.0001 & 0.69\% & 0.03\tabularnewline
		\hline 
		\multicolumn{1}{|c|}{ 3} & 1.5945 & 0.0028 & 1.4679 & 0.0001 & 7.94\% & 0.16\tabularnewline
		\hline 
	\end{tabular}
	\\~\\
	\end{center}
	$*$: Sum of emissions at every time step (i.e. time integration). Mean: average of 100 different measurements. $\sigma$: standard deviation.\\
	\label{table2}
\end{table}

{\noindent \bf Comment:} Note that the solution we have obtained is optimal for the environment and for the collective, e.g., in terms of the reduction of overall emissions. However, the solution might be unfair for some single users who would be recommended to drive at a different speed than originally desired. One way to improve fairness could be to decrease road taxes for virtuous vehicles, to compensate them from the inconvenience caused by the dirty vehicles in terms of recommended average speeds.\newline

{\noindent \bf Comment:} Note that it is clearly the case that the design parameters of the algorithm have the potential to affect emission savings. For example, the speed of convergence of the algorithm affects the rate of which the emissions are saved.\newline

\begin{figure}[h!]
\centering
\includegraphics[width=3.3in]{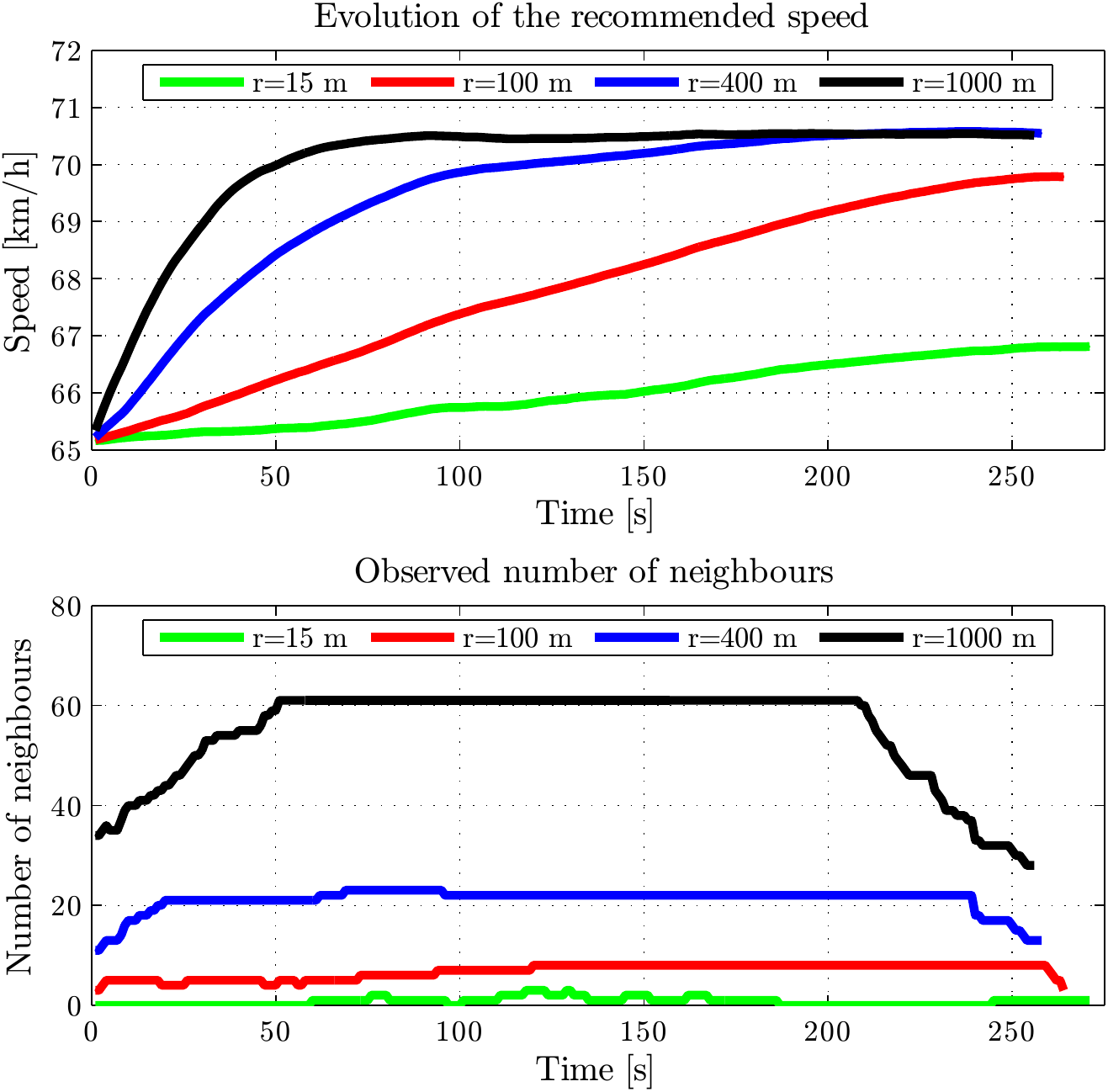}
\caption{Evolution of the recommended speed (top) and number of neighbours (bottom) of a target vehicle, as functions of the radius $r$ of its communication range.}
\label{Convergence_Issues}
\end{figure}

Following the previous comments, convergence issues are briefly discussed with the support of Fig. \ref{Convergence_Issues}. As we have mentioned in the previous paragraph, the recommended speed is updated in an iterative fashion, and it eventually converges to a steady-state value after a number of steps. The time of convergence is mainly affected by the topology of the communication graph and, in particular, by the number of neighbours. Convergence rate of algorithms of graphs is a well-studied topic. The interested readers are referred to \cite{olfati2007consensus,olshevsky2009convergence} for the favours of results available. In this paper, we just give a brief empirical illustration. As we shall see, roughly speaking, the larger is the communication range of the vehicles, the larger will be the number of neighbouring vehicles, and (on average) the earlier is the convergence to the steady-state value. In particular, Fig. \ref{Convergence_Issues} shows how convergence of a single vehicle is affected by its communication range. Note that convergence occurs after a few steps (less than three minutes) when the communication range is of the order of hundreds of meters. On the other side, convergence is very slow when the communication range is of only 15 meters, where it is frequent to get an empty set of neighbouring vehicles. Finally, observe that the steady-state value of the optimal speed is not constant as the optimal speed depends on the vehicles currently driving along the road section L2, and thus the types of vehicles (and in turn the optimal speed) change continuously.\newline

\begin{figure}[h!]
\centering
\includegraphics[width=3.3in]{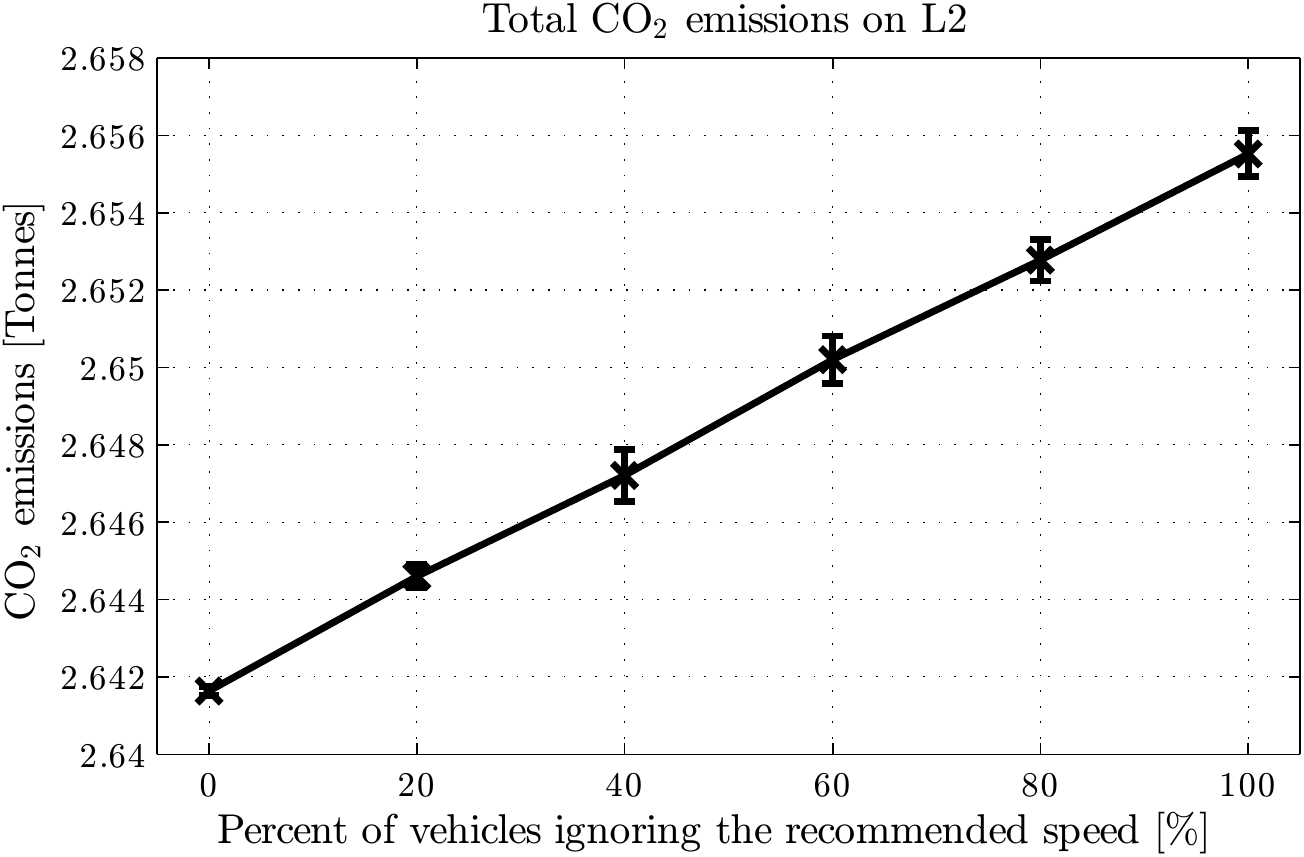}
\caption{Total CO$_2$ emissions as a function of the percentage of vehicles not following the recommendations: emissions are maximum when all vehicles do not follow the recommended speed, and minimum when all they do. Mean (x) and standard deviation (I) are calculated based on five different experiments.}
\label{cheating_vehicles}
\end{figure}

As has been previously remarked, convergence is for the recommended speeds, and thus it is independent on whether vehicles do follow the recommended speed or not. However, not following the recommended speed usually decreases the benefits of the proposed procedure in terms of CO$_2$ savings. Accordingly, Fig. \ref{cheating_vehicles} illustrates how CO$_2$ savings are affected by the percentage of vehicles that decide not to follow the recommended speed.

%-------------------------------------------------------------------------------------------------------------------------------------------------------------------------------------------
\subsection{SUMO-based hardware-in-the-loop (HIL) emulation}
%-------------------------------------------------------------------------------------------------------------------------------------------------------------------------------------------

\begin{figure}[h]
	\begin{center}
		{\includegraphics[width = 3.3in]{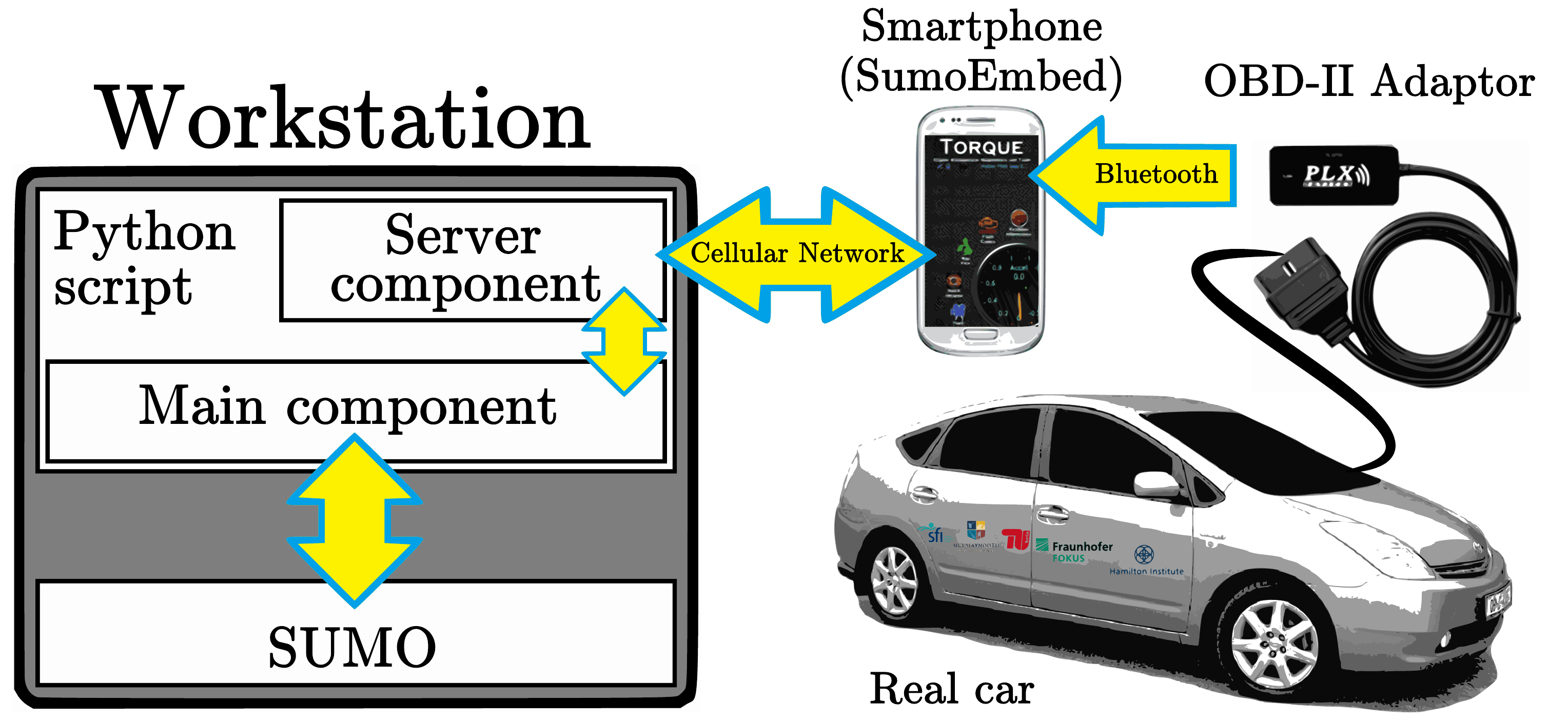}}
		\caption{Schematic of the SUMO-based HIL emulation platform.}
		\label{fig:Fig_HIL}
	\end{center}
\end{figure}

Finally, to give a feeling to a real driver of how this system might function, we now describe a HIL implementation of the algorithm. Specifically, we use a SUMO-based HIL
emulation platform that was developed at the Hamilton Institute \cite{Griggs2013,Griggs2014}. This emulation platform uses SUMO to simulate a real environment and generate virtual cars, along with a dedicated communication architecture supported by
TraCI (a Python script implementing a TPC-based client/server architecture) to provide on-line access to SUMO, a smartphone
connected to the 3G network and running the plug-in {\it SumoEmbed} (designed for use with Torque Pro \cite{Torque}),
and a OBD-II adaptor \cite{PLX} to embed a real car into the simulation, as shown
in Fig. \ref{fig:Fig_HIL}. The idea then is to allow a person to drive a real vehicle on a real street circuit, to experience being connected to a network of (virtual) vehicles driving along a virtual environment based on the physical street circuit. Specifically, we performed this experiment by driving a Toyota Prius on a single-lane street circuit in the North Campus of the Maynooth University, while the Prius is embedded into the HIL emulation and represented by an avatar which interacts with the avatars of 29 other virtual (simulated) vehicles driving along the same stretch of (emulated) road.

The experiment begins when the simulation is started on the workstation and the server component of the Python script waits
for a call from the OBD-II connected smartphone in the real vehicle. Since the selected street circuit only has one lane, the vehicles are released sequentially from the same starting point. The avatar representing the Prius departs in
the seventh position. Once the connection between the Prius
and the workstation is established, the position and speed of the Prius' avatar are updated using real-time information from
the Prius via the OBD-II adaptor. From the point of view of the ISA algorithm, the Prius is regarded as a normal agent in the SUMO simulation, i.e.
treated just like any other simulated vehicle.\newline

The consensus algorithm for the proposed ISA system is embedded in the main component Python script. Thus, once the 
respective recommended speeds are calculated, they are sent to the vehicles through the server component, via the cellular
network to the smartphone in the case of the Prius, and via TraCI commands in the case of the other vehicles in the simulation.
Note here that the driver behaviour is different for a simulated vehicle compared to the case of the Prius: while we force each
simulated vehicle to follow the recommended speed as far as possible\footnote{Concerning mainly the interaction between vehicles
	and the design parameters for the simulated cars such as acceleration, deceleration, car following model or driver information.},
the Prius' driver is allowed to either follow or ignore the speed recommendation (displayed on the smartphone's screen) as desired.\newline

The HIL experiment is setup as follows.

\begin{itemize}
\item Length of the experiment: 600 s, of which the ISA algorithm is only engaged at around time 300 s;
\item Total number of cars: 30, with uniform distribution of emission types between R014 and R040, and uniform distribution of vehicle types, with a maximum speed of 100 km/h.
\item The sampling time interval $\Delta T$ for collecting new information and updating the recommendations is 1 s. \newline
\end{itemize}

\begin{figure}[h]
	\begin{center}
		{\includegraphics[width = 3.3in]{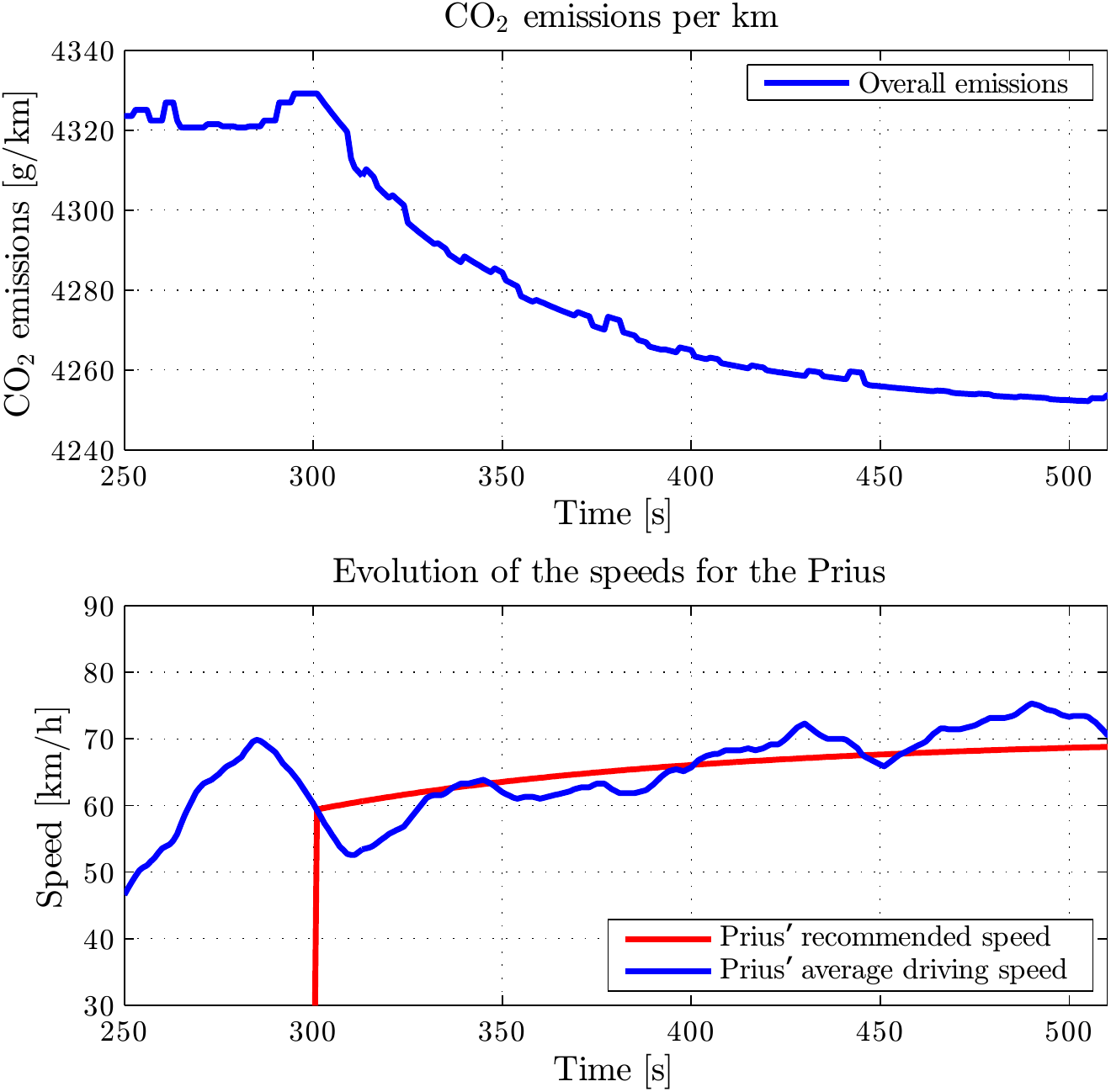}}
		\caption{Evolution of the overall CO$_2$ emissions (top) and of the speeds related with the Prius (bottom) for the HIL simulation. The algorithm was turned on around time 300 s. Average speed calculated with a window size of 20 time steps.}
		\label{fig:Fig_HIL_results1}
	\end{center}
\end{figure}

Results of the experiment are depicted in Fig.\ref{fig:Fig_HIL_results1}. From Fig.\ref{fig:Fig_HIL_results1} it can be observed that the total emissions oscillate around an average value of 4323 g/km once all the 30 vehicles are added to the simulation, and also that they start reducing once the ISA algorithm is switched on, up to a final average value of 4253 g/km.
Finally, as can be observed in  Fig.\ref{fig:Fig_HIL_results1} (bottom), the recommended speed can be roughly followed (on average) by the driver subject  to the physical constraints on the real street circuit (e.g. traffic calming devices), and the discretisation step for the visualisation of the recommended speed.

%%%%%%%%%%%%%%%%%%%%%%%%%%%%%%%%%%%%%%%%%%%%%%%%%%%
%%%%%%%%%%%%%%%%%%%%%%%%%%%%%%%%%%%%%%%%%%%%%%%%%%%
%%%%%%%%%%%%%%%%%%%%%%%%%%%%%%%%%%%%%%%%%%%%%%%%%%%
\section{A Speed Advisory System for Electric Vehicles}
%%%%%%%%%%%%%%%%%%%%%%%%%%%%%%%%%%%%%%%%%%%%%%%%%%%
%%%%%%%%%%%%%%%%%%%%%%%%%%%%%%%%%%%%%%%%%%%%%%%%%%%
%%%%%%%%%%%%%%%%%%%%%%%%%%%%%%%%%%%%%%%%%%%%%%%%%%%

In this section, we extend our previous discussion to the special case of a fleet of EVs. The motivation for doing so is the recent interest in EVs as a cleaner alternative to conventional, more polluting, ICE vehicles. In recent years, some cities have decided to close the city centres to normal traffic, only allowing the transit to specific categories of low (or zero) polluting vehicles; see for instance \cite{umweltzonen}. Similarly, in many cities all around the world, the urban public transportation fleet has been restricted to electrically driven vehicles, as for the case of fleets of urban electric buses; see for instance the cases of S\~{a}o Paulo \cite{cities-today} in Latin America, Louisville in the US \cite{cleantechnica}, or Wien in Europe \cite{siemens}.\newline

Clearly, the previously designed Algorithm I needs to be adapted when applied to a fleet of EVs concerning different aspects. First, in terms of the field of application, EVs are typically used for short distances due to their reduced driving ranges, and thus are most likely deployed in city centres; second, the cost functions should not consider instantaneous emissions that, in the case of EVs, can be considered equal to zero. Accordingly, in the following we show how the previous framework can be adapted to the case of a fleet of EVs, and now the objective is to maximise the energy efficiency of the fleet of cars or, in other words, to extend their driving range.

%-------------------------------------------------------------------------------------------------------------------------------------------------------------------------------------------
\subsection{Cost functions to represent energy consumption in EVs}
%-------------------------------------------------------------------------------------------------------------------------------------------------------------------------------------------

Most of the discussion here follows the reference \cite{vanHaaren}, where the ranges of EVs are reported for different brands and under different driving cycles. Power consumption in an EV driving at a steady-state speed (along a flat road) is caused mainly by four sources:
\begin{itemize}
	\item {\bf Aerodynamics power losses}: they are proportional to the cube of the speed of the EV, and depend on other parameters typical of a single vehicle such as its frontal area and the drag coefficient (which in turn depends on the shape of the vehicle).
	\item {\bf Drivetrain losses}:  they result from the process of converting energy in the battery into torque at the wheels of the car. Their computation is not simple, as losses might occur at different levels (in the inverter, in the induction motor, gears, etc); in some cases, these power losses have been modelled as a third-order polynomial, whose parameters have been obtained by fitting some experimental data (see \cite{vanHaaren}).
	\item {\bf Tires}: the power required to overcome the rolling distance depends on the weight of the vehicle (and thus, on the number of passengers as well), and is proportional to the speed of the vehicle.
	\item {\bf Ancillary systems}: this category includes all other electrical loads in the vehicle, such as HVAC systems, external lights, audio system, battery cooling systems, etc. Here, the power consumption does not depend on the speed of the vehicle and can be represented by a constant term that depends on external factors (e.g., weather conditions) and personal choices (desired indoor temperature, volume of the radio, etc). According to experimental evaluations \cite{vanHaaren}, the power losses due to ancillary services usually vary between $0.2$ and $2.2$ kW.\newline
\end{itemize}

Thus, by summing up all the previous terms, the power consumption $P_{cons}$ can be represented as a function of the speed $v$ as
\begin{equation}
	\frac{P_{cons}}{v} = \frac{\alpha_0}{v} + \alpha_1 + \alpha_2 v + \alpha_3 v^2,
	\label{power_consumption}
\end{equation}
where the left hand side is divided by the speed in order to obtain an indication of energy consumption per km, expressed in kWh/km. Such a unit of measurement is usually employed in energy-efficiency evaluations, and we shall assume that every single EV will use \eqref{power_consumption} as its personal cost function $f_i$.
Accordingly, Fig. \ref{Relationship_Example} shows a possible relationship between speed and power consumption, obtained using data from Tesla Roadster and assuming a low power consumption for ancillary services of 0.56 kW (i.e., assuming air conditioning switched off).
As can be noted from  Fig. \ref{Relationship_Example}, there is a large energy consumption at large speeds due to the fact that power increases with the cube of the speed for aerodynamic reasons; however, it is also large for low speeds, due to the fact that travel times increase and, accordingly, constant power required by ancillary services demands more energy than the same services delivered with high speeds.

%-------------------------------------------------------------------------------------------------------------------------------------------------------------------------------------------
\subsection{Experimental results}
%-------------------------------------------------------------------------------------------------------------------------------------------------------------------------------------------

According to the previous discussion, we now assume that the objective is to infer the optimal speed that the ISA system should broadcast to a fleet of EVs travelling in a given area of a city (e.g., in the city centre). For this purpose, we assume that a fleet of $100$ vehicles travels in the city centre for an hour, and following the next steps:
\begin{itemize}
\item In the first 20 minutes, the vehicles travel at the optimal speed calculated from Algorithm 1.
\item In the second 20 minutes, they travel at a speed below the optimal speed.
\item In the last 20 minutes, they travel at a speed above the optimal speed.\newline
\end{itemize}

In the first stage we assume that the communication graph among the EVs changes in a random way, i.e., at each time step an EV receives information from a subset of vehicles belonging to the fleet. This is a simplifying assumption that can be justified by assuming that in principle all vehicles might communicate to all the other vehicles (i.e., they are relatively close), but some communications might fail due to obstacles, shadowing effects, external noise, etc. Besides, in the two last stages we assume that the change of speed occurs almost instantaneously, since there is no requirement to iteratively compute an optimal speed. \newline

\begin{figure}[ht]
\centering
\subfloat[An individual cost function. ]{\includegraphics[width=3.3in]{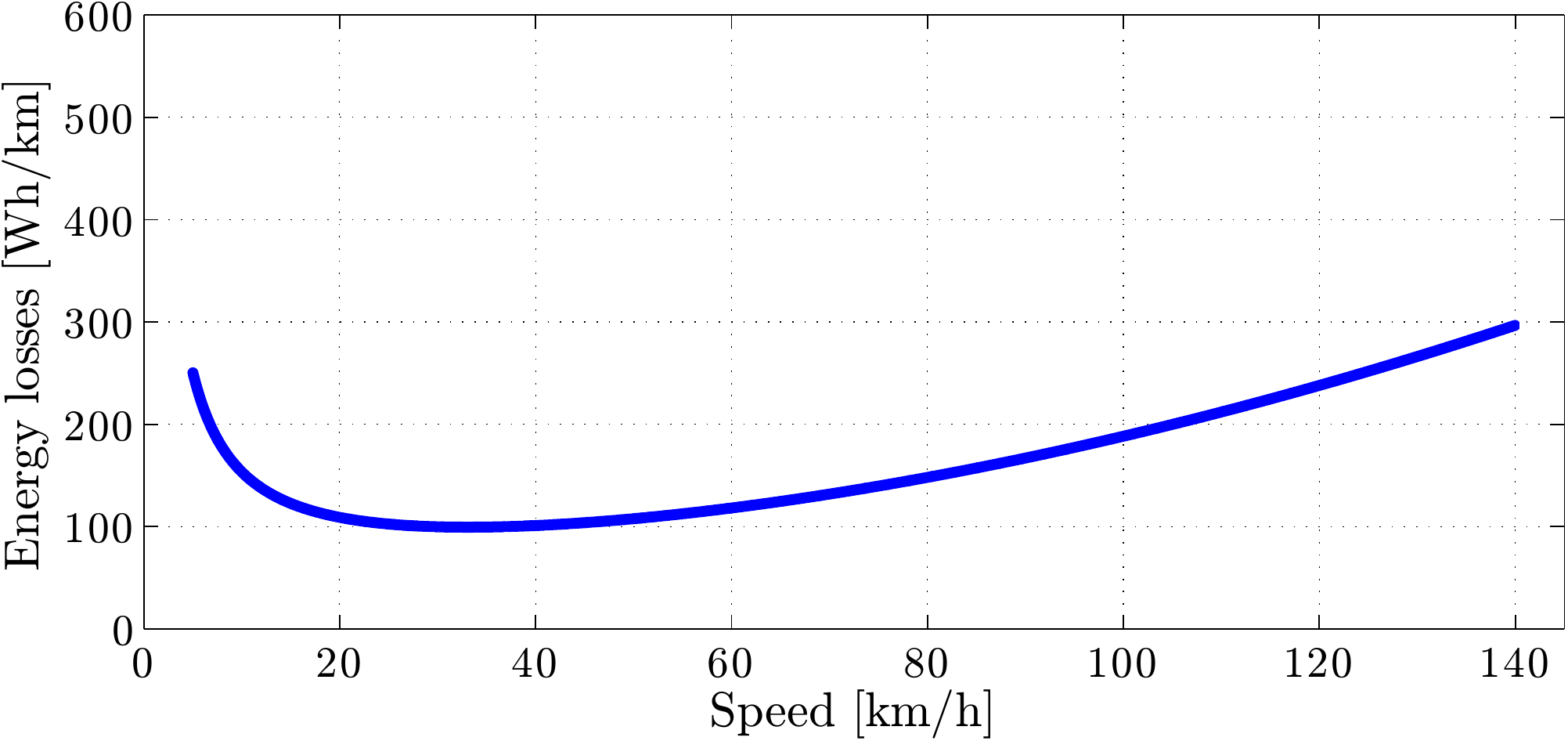}
\label{Relationship_Example}}
\hfil
\\
\subfloat[All the cost functions overlapped. Black circles mark the minimum point on each curve. ]{\includegraphics[width=3.3in]{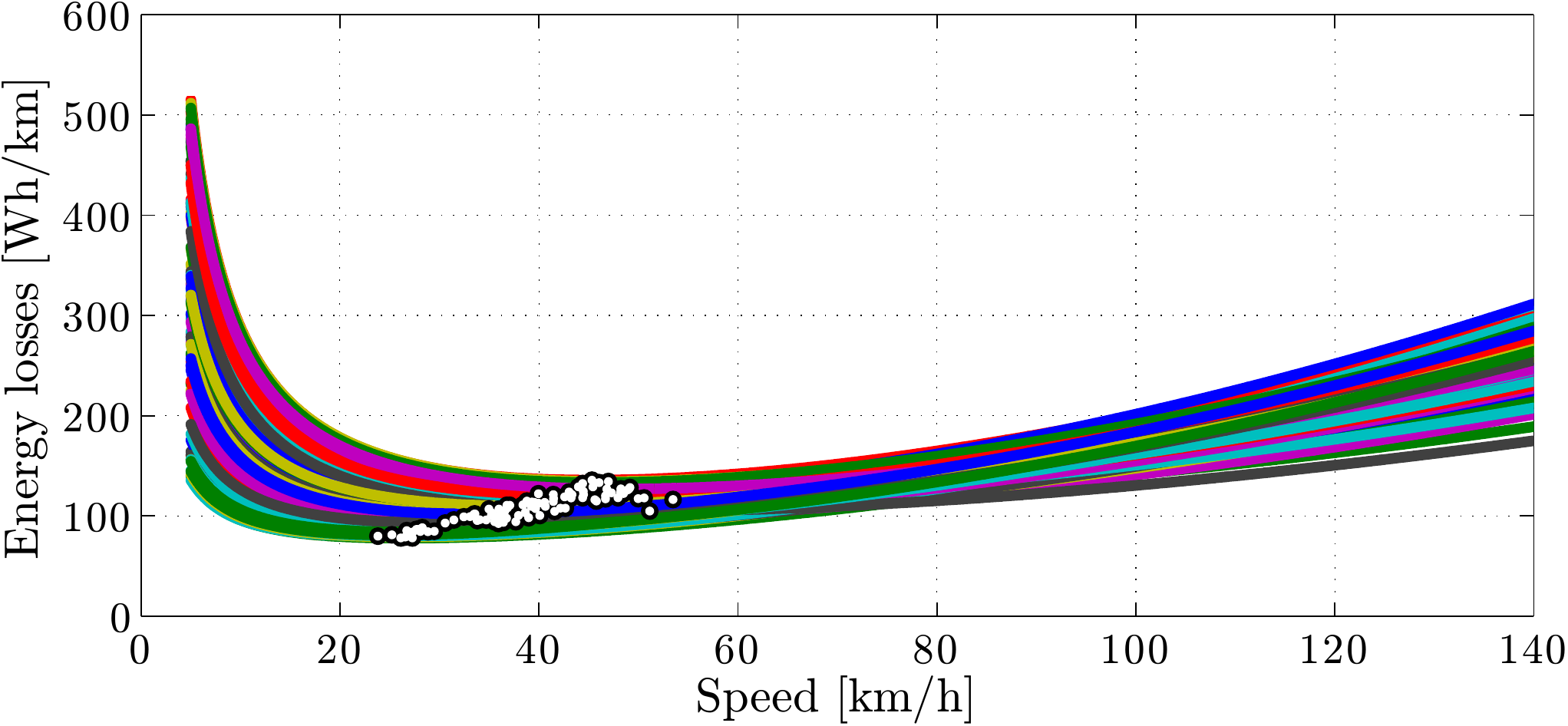}
\label{Utility_Functions}}
\caption{Curves for the cost functions used in the experiment. All of them were chosen convex.}
\label{fig:fig10}
\end{figure}

\begin{figure}[ht]
\centering
\subfloat[Evolution of the vehicles' speeds.]{\includegraphics[width=3.3in]{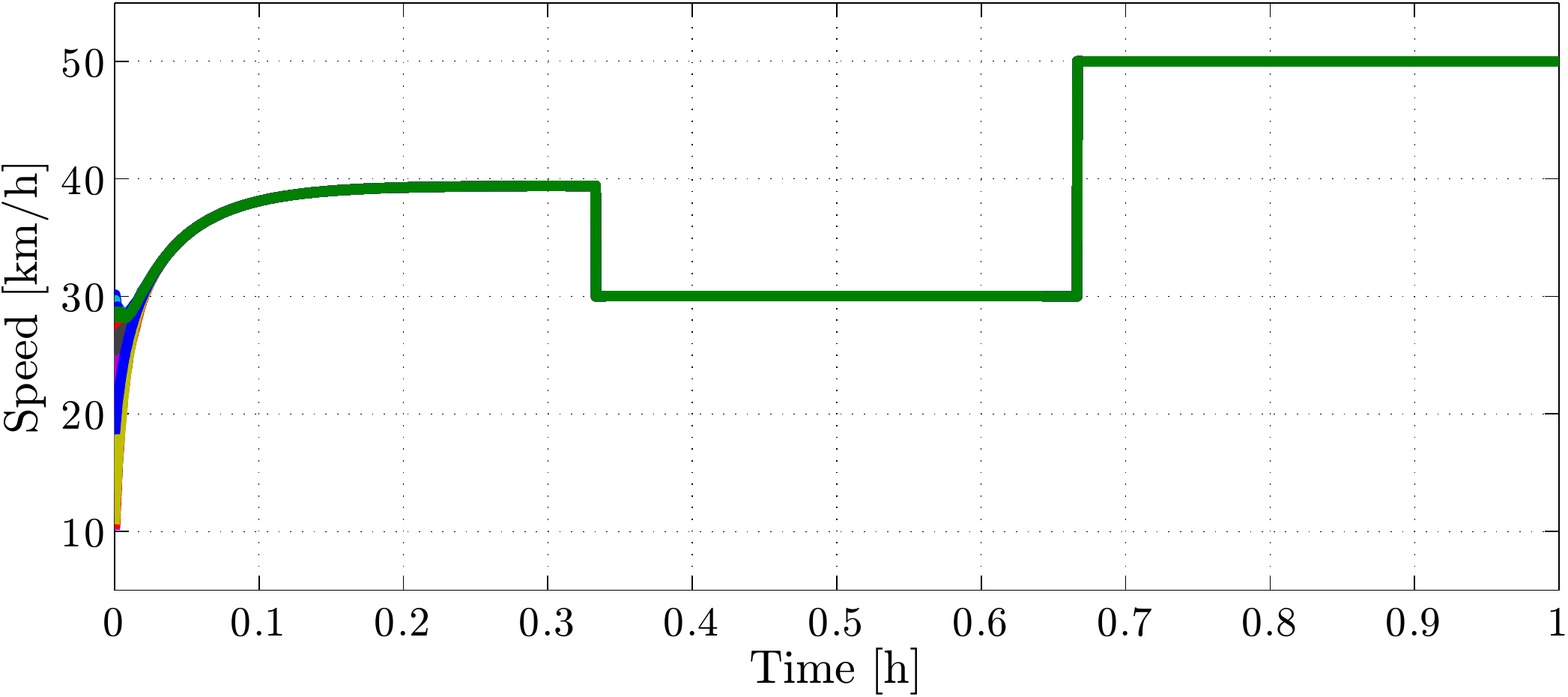}
\label{Sim2_Speed}}
\hfil
\\
\subfloat[Evolution of the overall energy loss.]{\includegraphics[width=3.3in]{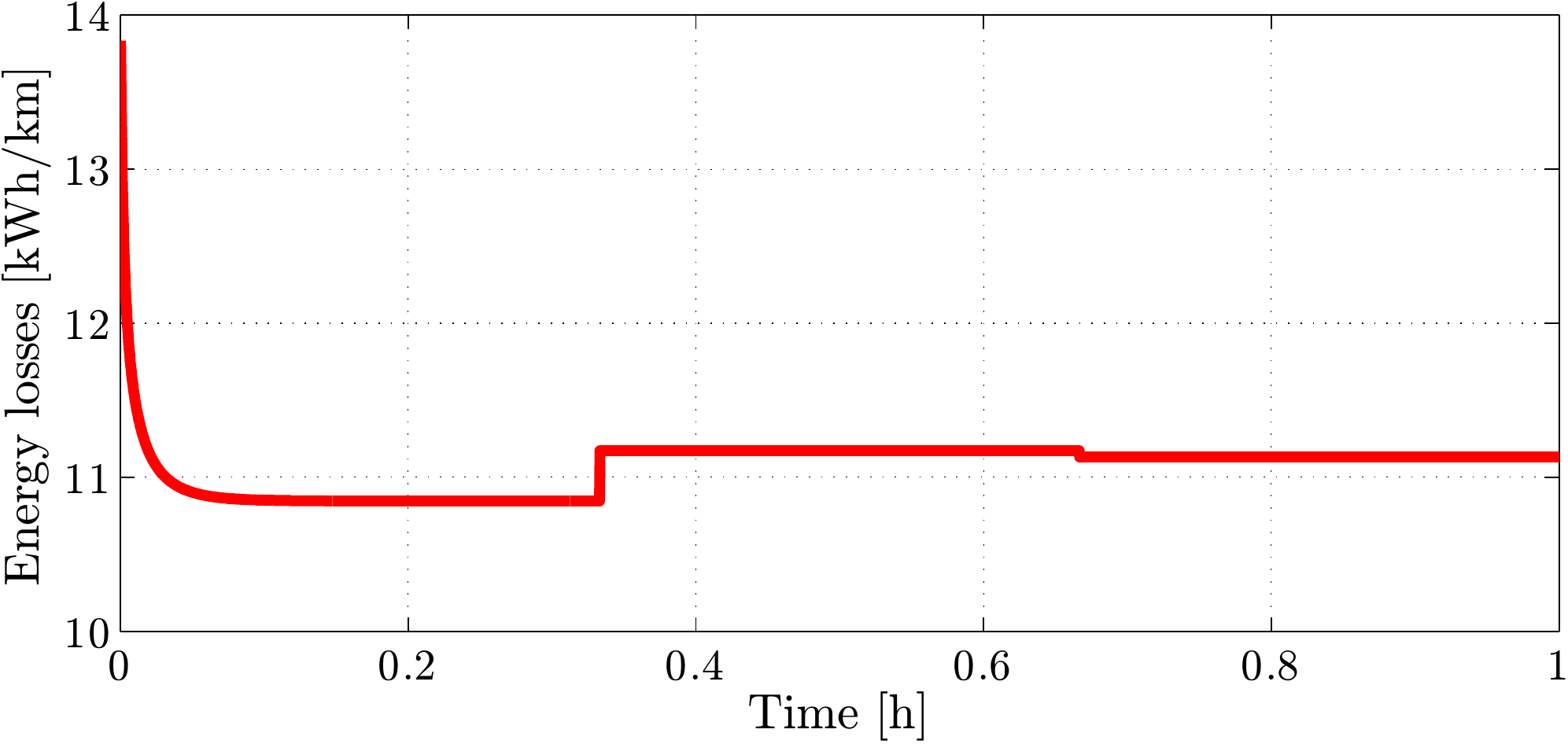}
\label{Sim2_Energy}}
\label{fig:fig11}
\caption{Simulation results for the network of EVs: Algorithm 1 is applied until time 0.33 h, and then two different speeds (below and above the optimal one) are suggested in $[0.33,0.66]$ h and $[0.66,1]$ h, respectively.}
\end{figure}

We tuned our parameters in Algorithm \ref{alg:Algorithm1} as $\eta = \mu = 0.001$, and we simulate different cost functions for each EVs by assuming a random number of people inside each car (between 1 and 5 people) with an average weight of 80 kg, and by assuming a different consumption from ancillary services within the typical range of $[0.2, 2.2]$ kW. The curves of the cost functions used in our experiment are shown in Fig. \ref{Utility_Functions}. The evolution of the speeds of the EVs are shown in Fig. \ref{Sim2_Speed}, while the average energy consumption is shown in Fig. \ref{Sim2_Energy}.

As is shown in Fig. \ref{Sim2_Energy}, a simple ISA can be used for the case of a fleet of EVs just as in the case of conventional cars (for what regards the mathematical background) by simply adapting the cost functions to the new application of interest.

%%%%%%%%%%%%%%%%%%%%%%%%%%%%%%%%%%%%%%%%%%%%%%%%%%%
%%%%%%%%%%%%%%%%%%%%%%%%%%%%%%%%%%%%%%%%%%%%%%%%%%%
%%%%%%%%%%%%%%%%%%%%%%%%%%%%%%%%%%%%%%%%%%%%%%%%%%%
\section{Conclusion}  \label{Conclusion}
%%%%%%%%%%%%%%%%%%%%%%%%%%%%%%%%%%%%%%%%%%%%%%%%%%%
%%%%%%%%%%%%%%%%%%%%%%%%%%%%%%%%%%%%%%%%%%%%%%%%%%%
%%%%%%%%%%%%%%%%%%%%%%%%%%%%%%%%%%%%%%%%%%%%%%%%%%%

In this paper we present a new ISA system. The system is based on a solution to an optimised consensus problem. We show that the ISA can be implemented in a privacy preserving fashion, in a manner that accounts for vehicle density and composition, and in a manner that is provably convergent. Simulations are given to illustrate the efficacy and the acceptability of the algorithm. Finally, the algorithm has been implemented in a real production vehicle embedded into a HIL emulation using nothing more than a smartphone and a commercially available OBD-II plug-in. In future work we intend to extend the HIL simulation to include more vehicles (and thus, more drivers) to better test the compliance of people to follow the recommended speed.

%-------------------------------------------------------------------------------------------------------------------------------------------------------------------------------------------
\appendix[Outline of proof  of Theorem \ref{thm:thm01}] \label{Proofoutline}
%-------------------------------------------------------------------------------------------------------------------------------------------------------------------------------------------

  In this section, we give an outline of the proof for the claims in
  Subsection \ref{s:Problem_Statement}, in which we largely rely on the
  results obtained in \cite{TAC}.\newline

  In Theorem \ref{thm:thm01}, statement (i) is a consequence of the
  Banach contraction theorem. It is a straightforward calculation to show
  that the bounds \eqref{rangemu} ensure that the function $h$ defining
  the Lure system \eqref{eq:fw-tvsys0_1dim} is in fact a global strict
  contraction on $\R$. Statement (ii) then follows directly from the
  definition of $h$: if $h(y^*) = y^*$ then $G(y^* e)=0$ and by
  \eqref{eq:Gdef} this is equivalent to the optimality condition
  \eqref{optcondition}. The global optimality of $y^* e$ for the
  optimisation problem \eqref{eq:opt} of this optimal point
  then follows as \eqref{optcondition} is the standard first order
  necessary condition for optimality, and because strict convexity of the
  cost functions implies this condition is also
  sufficient. Uniqueness is a further consequence of strict optimality.\newline

It therefore remains to show that statement \ref{thm:thm01} (iii) holds. To
this end we recall the following two lemmas from \cite{TAC}.

\begin{lemma}[\cite{TAC}]\label{lem:solspane}
	Let $\left\{ P\left(k\right) \right\}_{k\in\N} $ be a sequence of row-stochastic matrices.
	If $\left\{ y\left(k\right) \right\}_{k\in\N}$ is a solution of
        the Lure system \eqref{eq:fw-tvsys0_1dim} then $\left\{ y\left(k\right) e\right\}_{k\in\N}$ is a solution of \eqref{eq:fw-tvsys0}.
\end{lemma}

\begin{lemma}[\cite{TAC}]\label{lem:boundedsols}
	Let $\left\{ P\left(k\right) \right\}_{k\in\N}$ be a  strongly ergodic sequence of row-stochastic matrices,
	and suppose that $G: \R^n\rightarrow \R$ is continuous and satisfies the following conditions:
	\begin{enum} 
		\item there exists an $\varepsilon>0$
		such that $G$ satisfies a Lipschitz condition with constant $L>0$ on
		the set
		\begin{equation*}
		B_\varepsilon\left(E\right) := \left\{ x \in \R^n \,:\, \dist\left(x,E\right) \leq \varepsilon \right\},
		\end{equation*} 
		where $\dist\left(x,E\right):= \inf \left\{ \left\Vert x - z \right\Vert: z \in E \right\}$ is the distance of a vector $x\in \R^n$
		to the consensus set $E:= \spann\left\{ e \right\}$; and
		
		\item there exists constants $\beta,\gamma > 0$ such that
		\begin{equation*}
		\left| h\left(y\right) \right| \leq  \left|y\right| - \gamma\,\,\,\, \mbox{when} \,\,\,\, \left|y\right| \geq \beta,
		\end{equation*}
		where $h\left(y\right) = y + G\left(ye\right)$.
	\end{enum}
	Then, every trajectory of \eqref{eq:fw-tvsys0} is bounded.
\end{lemma}

It is easy to see that $G$ as defined in \eqref{eq:Gdef} satisifies the
conditions of Lemma \ref{lem:boundedsols}. Indeed, $G$ is even globally
Lipschitz continuous because of the Lipschitz continuity assumption \eqref{bdd}.
Furthermore, as $h$ is a strict contraction on $\R$ with fixed point $y^*$,
we may denote the contraction constant of $h$ by $0<c<1$ and obtain for
any $y \in \R$ such that
\begin{align*}
\left| h\left(y\right) \right| &\leq \left| h\left(y\right) \footnotesize{-} y^{*} \right| \footnotesize{+}  \left| y^{*}\right|
&& \leq  c \left| y \footnotesize{-} y^{*} \right| \footnotesize{+}  \left| y^{*} \right|\\
&\leq c \left| y \right| \footnotesize{+} \left(1\footnotesize{+}c\right) \left| y^{*} \right|
&&= \left| y \right| \footnotesize{-} \left(1\footnotesize{-}c\right) \left| y \right| \footnotesize{+}   \left(1\footnotesize{+}c\right) \left| y^{*} \right|,
\label{hieq}
\end{align*}
from which it is easy to derive constants $\beta$ and $\gamma$.\newline

Finally, if every trajectory of \eqref{eq:fw-tvsys0} is bounded, then
every trajectory has a nonempty bounded $\omega$-limit set. Because of the
averaging property of stochastic matrices and the assumption of uniform
strong ergodicity, this $\omega$-limit set is a subset of the span of $e$.
By part (i) of the theorem, the Lure system has a globally asymptotically
stable fixed point. Lemma \ref{lem:solspane} on the other hand ensures
that on $\mathrm{span}\, \{ e \}$ the trajectories of \eqref{eq:fw-tvsys0}
and \eqref{eq:fw-tvsys0_1dim} (multiplied by $e$) coincide.  It follows
that restricted to $\mathrm{span}\, \{ e \}$, the optimisation algorithm
\eqref{eq:fw-tvsys0} has only one $\omega$-limit set, namely $y^* e$.  It
then follows from a continuity argument that $y^* e$ is a globally
asymptotically stable fixed point of \eqref{eq:fw-tvsys0}.

\bibliographystyle{ieeetran}
\bibliography{References}

\vspace{-0.5cm}
\begin{IEEEbiography}[{\includegraphics[width=1in,height=1.25in,clip,keepaspectratio]{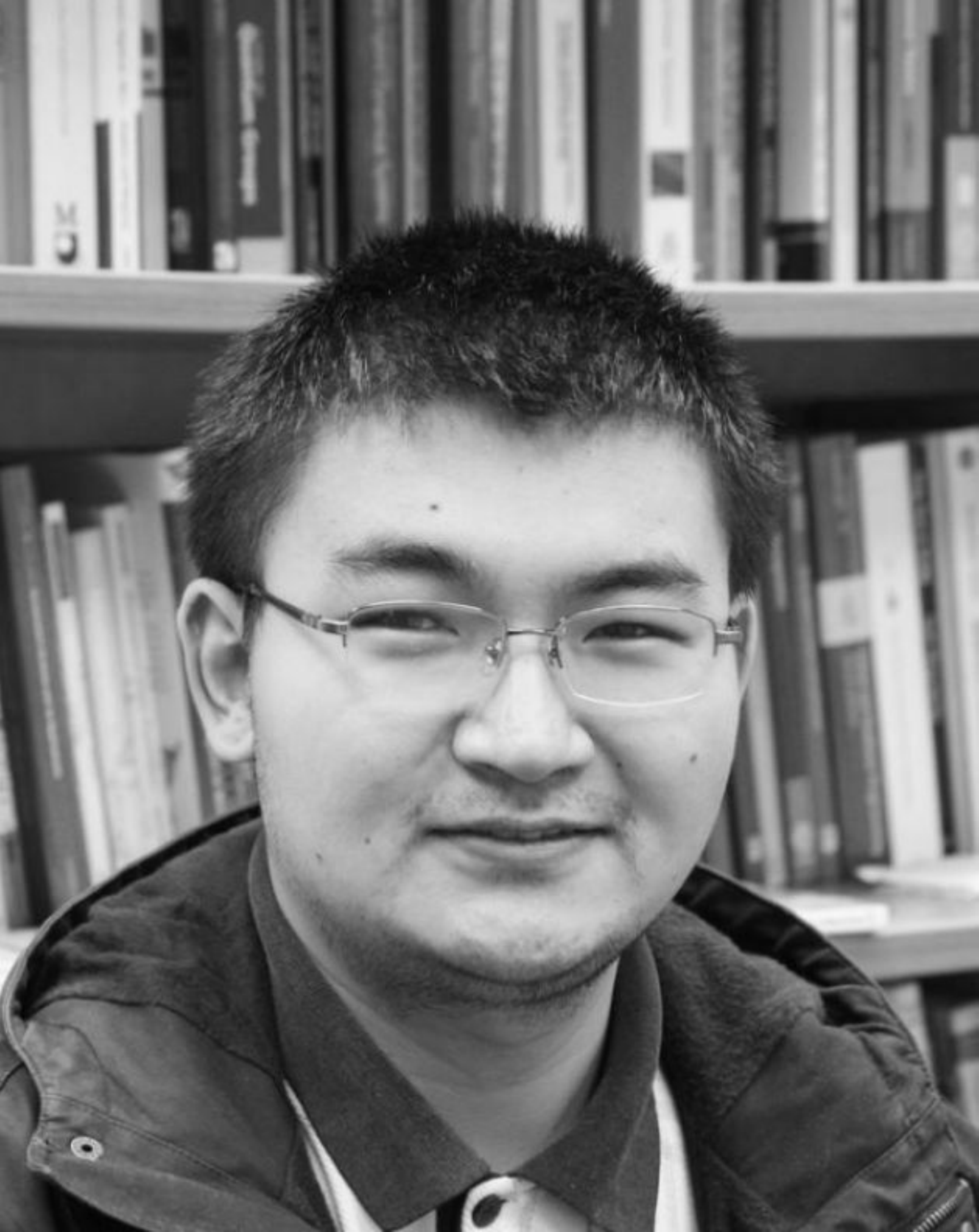}}]{Mingming~Liu}
received his double B.E. degrees in Electronic Engineering with first class honours from National University of Ireland Maynooth (now Maynooth University) and Changzhou University in 2011. He is currently working towards a Ph.D. degree at the Hamilton Institute, Maynooth University and he has also held a research assistant (pending viva) position with University College Dublin, working with Prof. R. Shorten, since June 2015. His current research interests are nonlinear system dynamics, distributed control techniques, modelling and optimisation in the context of smart grid and smart transportation systems. 		
\end{IEEEbiography}

\begin{IEEEbiography}[{\includegraphics[width=1in,height=1.25in,clip,keepaspectratio]{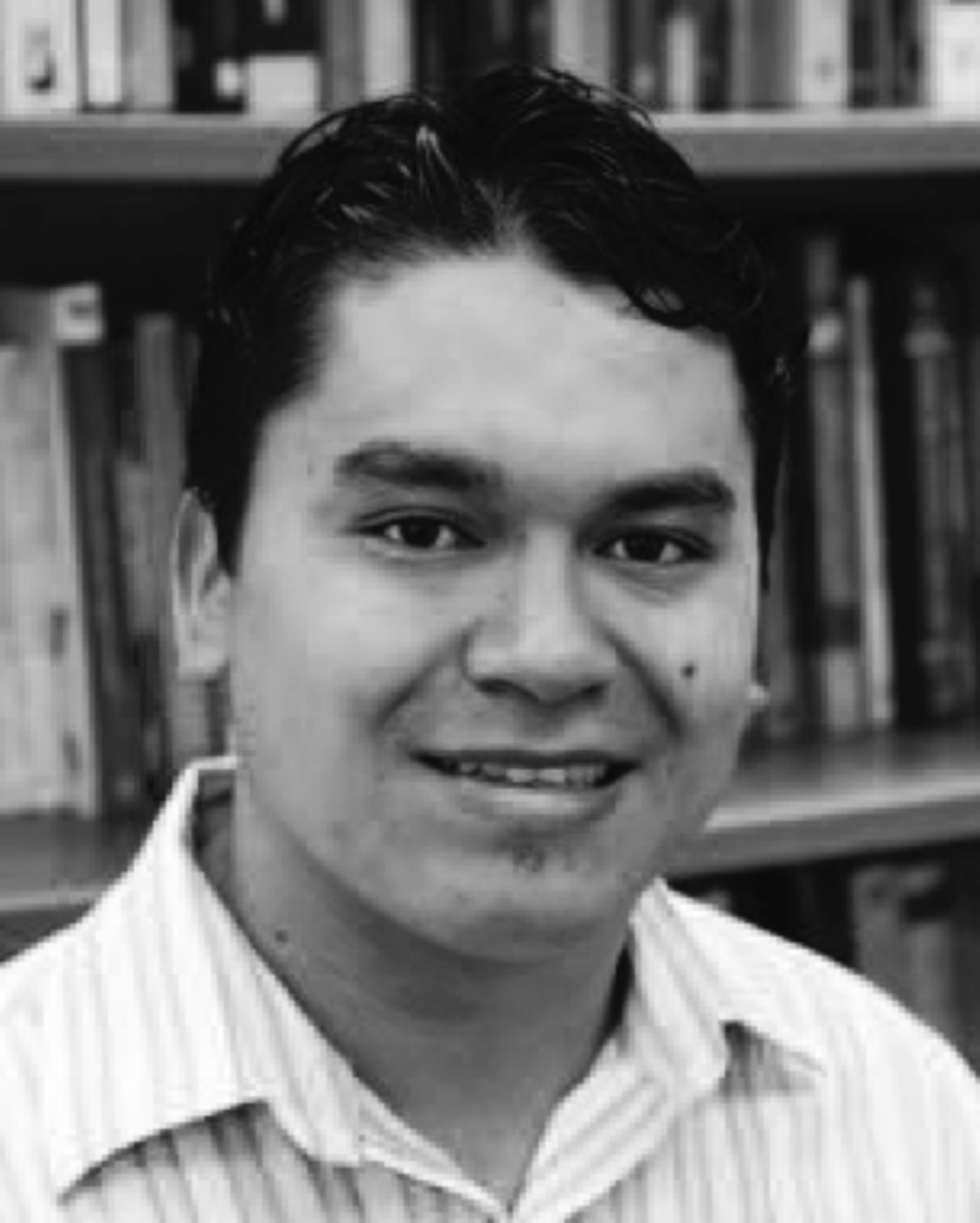}}]{Rodrigo H. Ord\'{o}\~{n}ez-Hurtado}
received his degree in Engineering in Industrial Automatica from the University of Cauca, Colombia, in 2005. He commenced his Ph.D. in Electrical Engineering at the University of Chile in March 2008, and between April-November 2012, he was an intern at the Hamilton Institute, a multi-disciplinary research centre established at the Maynooth University. In November 2012, he received his Ph.D. degree from the University of Chile, and between December 2012 - May 2015 he had a full-time postdoctoral position at the Hamilton Institute working with Professor R. Shorten and his research group. Currently, he has held a part-time postdoctoral position in both Maynooth Unversity and University College Dublin since June 2015, also working with Professor R. Shorten and his research group.  Rodrigo's interests include robust adaptive systems (control and identification), stability of switched systems, swarm intelligence, large-scale systems and intelligent transportation systems. His focus is on applications to the mining industry and transportation systems.
\end{IEEEbiography}

\begin{IEEEbiography}[{\includegraphics[width=1in,height=1.25in,clip,keepaspectratio]{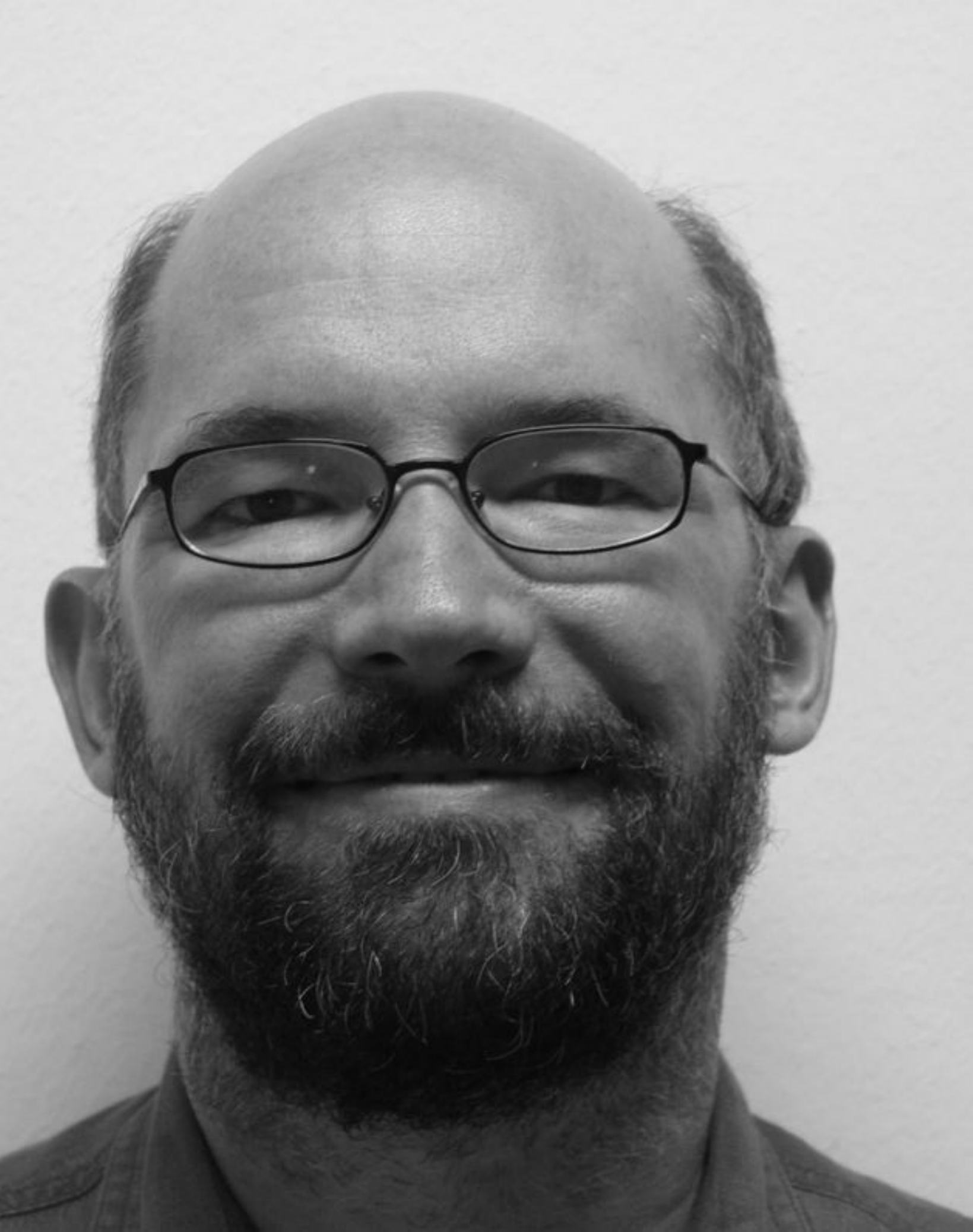}}]{Fabian~Wirth}
received his PhD from the Institute of Dynamical Systems at the University of Bremen in 1995. He has since held positions in Bremen, at the Centre Automatique et Syst{\`e}mes of Ecole des Mines, the Hamilton Institute at NUI Maynooth, Ireland, the University of W\"{u}rzburg and IBM Research Ireland. He now holds the chair for Dynamical Systems at the University
of Passau in Germany. His current interests include stability theory, queueing networks, switched systems and large scale networks with applications to networked systems and in the domain of smart cities.
	
\end{IEEEbiography}

\begin{IEEEbiography}[{\includegraphics[width=1in,height=1.25in,clip,keepaspectratio]{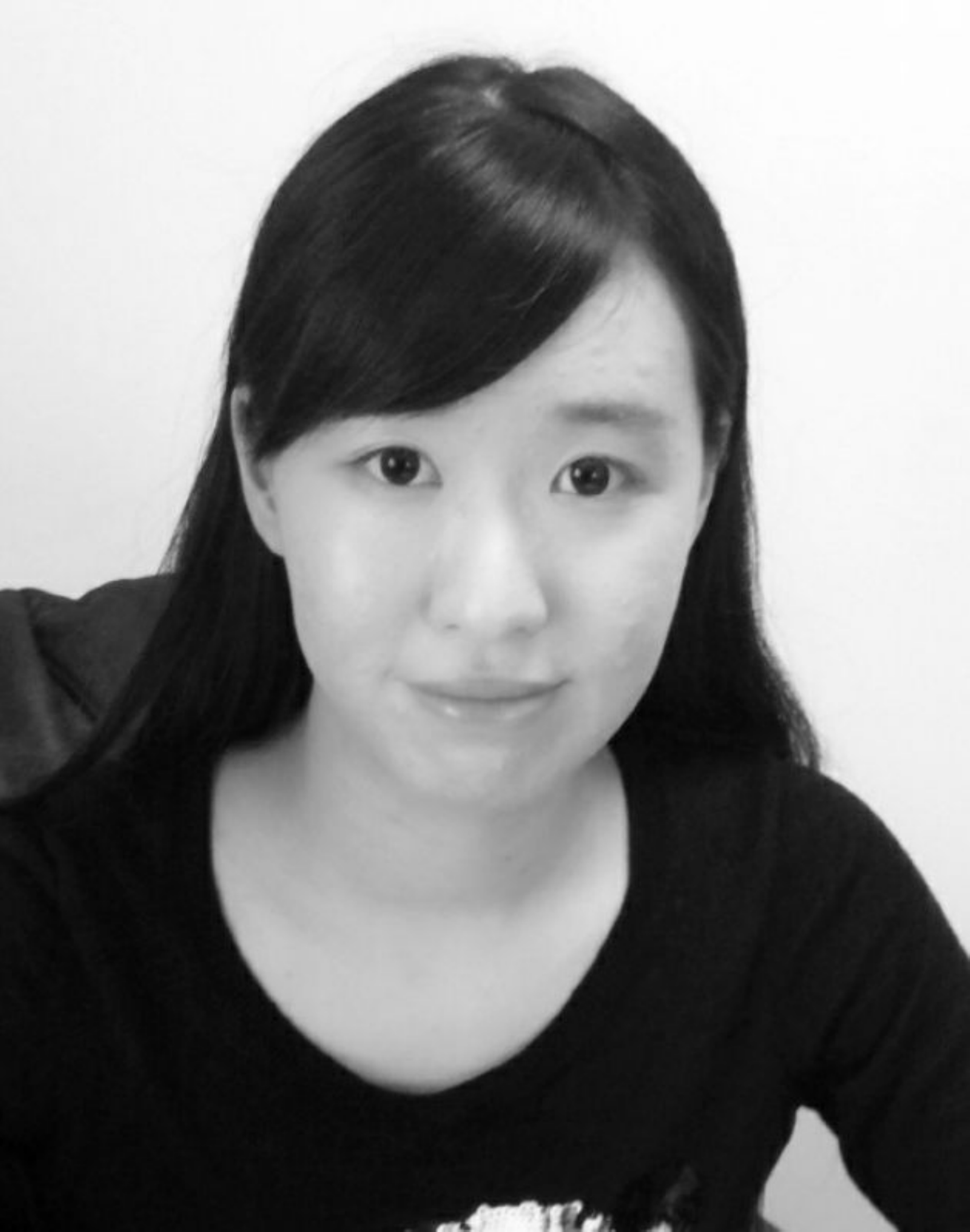}}]{Yingqi Gu}
received her B.E. degree in Electronic Engineering (first class honours) from Maynooth University in 2013. She obtained her M.Sc degree in Signal Processing and Communications at School of Engineering, University of Edinburgh in 2014. From February 2015, she commenced her Ph.D. degree in the University College Dublin with Prof. Robert Shorten. Her current research interests are modelling, simulation and optimisation of intelligent transportation systems.   		
\end{IEEEbiography}

\begin{IEEEbiography}[{\includegraphics[width=1in,height=1.25in,clip,keepaspectratio]{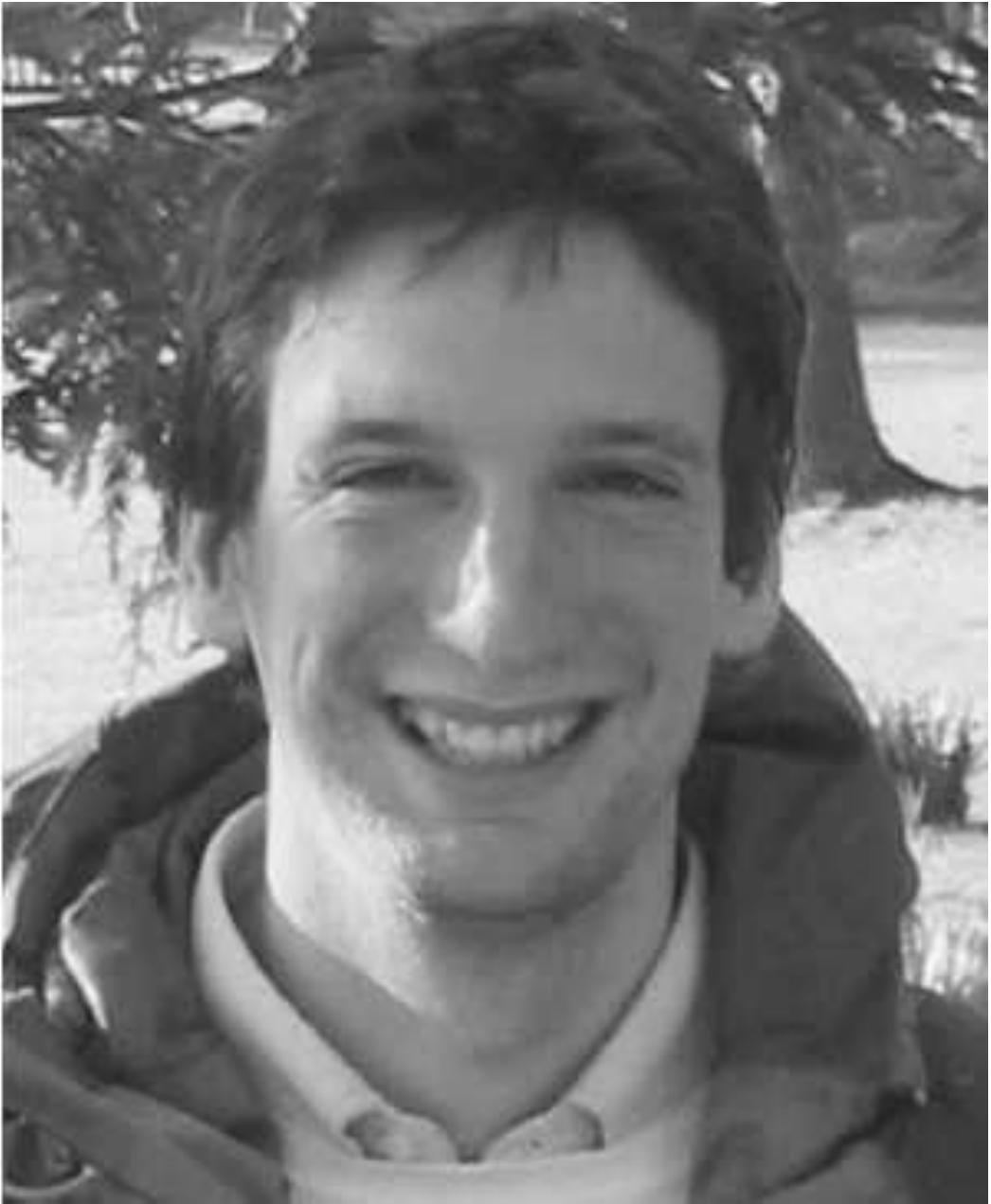}}]{Emanuele Crisostomi}
received the B.S. degree in computer science engineering, the M.S. degree in automatic control, and the Ph.D. degree in automatics, robotics, and bioengineering, from the University of Pisa, Italy, in 2002, 2005, and 2009, respectively. He is currently an Assistant Professor of electrical engineering with the Department of Energy, Systems, Territory and Constructions Engineering, University of Pisa. His research interests include control and optimization of large-scale systems, with applications to smart grids and green mobility networks.
\end{IEEEbiography}

\begin{IEEEbiography}[{\includegraphics[width=1in,height=1.25in,clip,keepaspectratio]{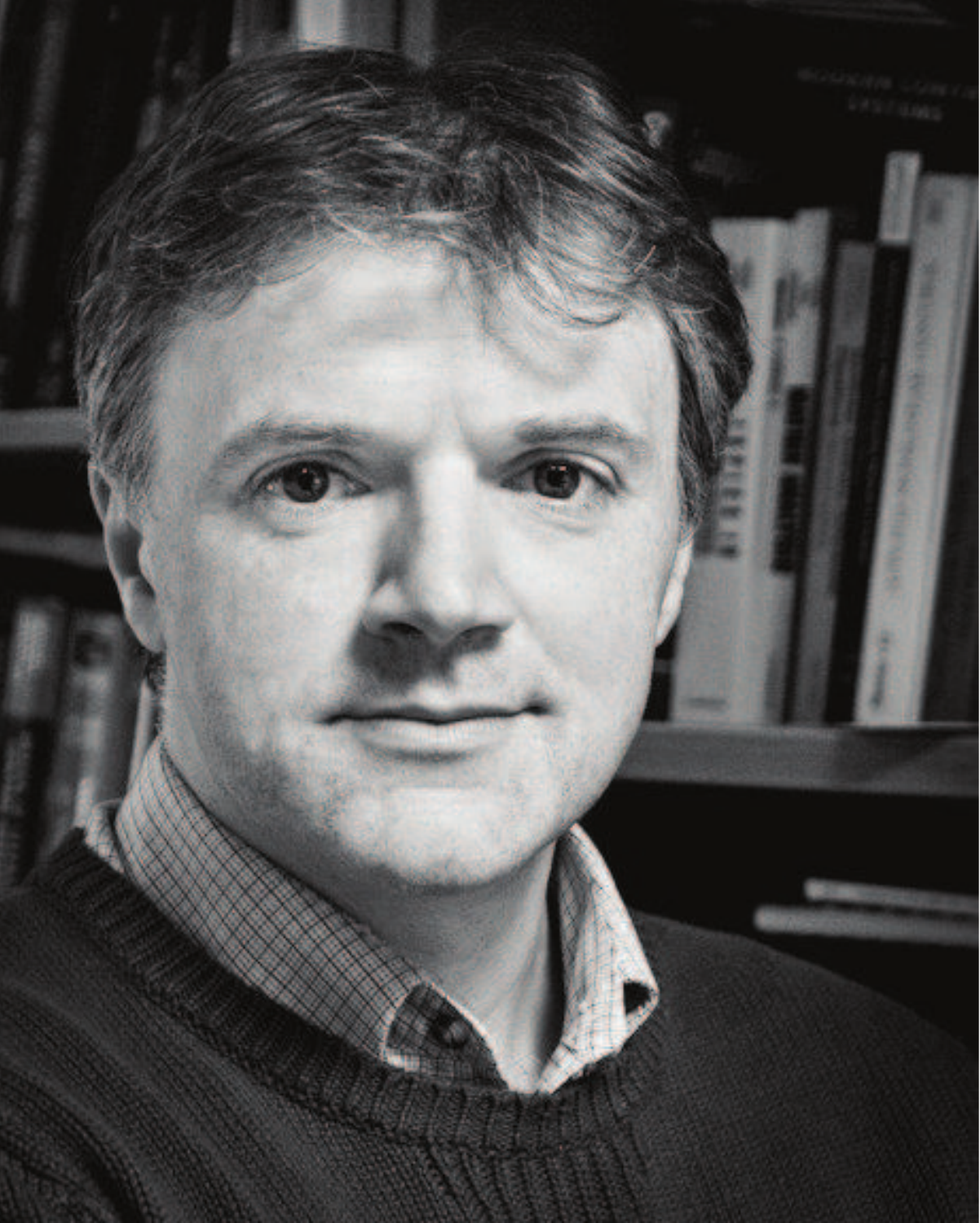}}]{Robert Shorten}
graduated from University College Dublin (UCD) in 1990 with a First Class Honours B.E. degree in Electronic Engineering. He was awarded a Ph.D. in 1996, also from UCD, while based at Daimler-Benz Research in Berlin, Germany. From 1993 to 1996, Prof. Shorten was the holder of a Marie Curie Fellowship at Daimler-Benz Research to conduct research in the area of smart gearbox systems. Following a brief spell at the Center for Systems Science, Yale University, working with Professor K. S. Narendra, Prof. Shorten returned to Ireland as the holder of a European Presidency Fellowship in 1997. Prof. Shorten is a co-founder of the Hamilton Institute at NUI Maynooth, where he was a full Professor until March 2013. He was also a Visiting Professor at TU Berlin from 2011-2012. Professor Shorten is currently a senior research manager at IBM Research Ireland. Prof. Shorten's research spans a number of areas. He has been active in computer networking, automotive research, collaborative mobility (including smart transportation and electric vehicles), as well as basic control theory and linear algebra. His main field of theoretical research has been the study of hybrid dynamical systems.
\end{IEEEbiography}

\end{document}